\documentclass[twocolumn]{aastex631}
\usepackage{threeparttable}
\usepackage{amsmath}

\begin{document}
%\linenumbers

\title{RRAT~J1913$+$1330: an extremely variable and puzzling pulsar}

\correspondingauthor{X. F. Wu, P. Jiang, Z. G. Dai, B. Zhang}
\email{xfwu@pmo.ac.cn, pjiang@nao.cas.cn, daizg@ustc.edu.cn, zhang@physics.unlv.edu}

\author{S. B. Zhang}
\affiliation{Purple Mountain Observatory, Chinese Academy of Sciences, Nanjing 210023, China}

\author{J. J. Geng}
\affiliation{Purple Mountain Observatory, Chinese Academy of Sciences, Nanjing 210023, China}

\author{J. S. Wang}
\affiliation{Max-Planck-Institut f\"ur Kernphysik, Saupfercheckweg 1, D-69117 Heidelberg, Germany}

\author{X. Yang}
\affiliation{Purple Mountain Observatory, Chinese Academy of Sciences, Nanjing 210023, China}
\affiliation{School of Astronomy and Space Sciences, University of Science and Technology of China, Hefei 230026, China}

\author{ J. Kaczmarek}
\affiliation{CSIRO Space and Astronomy, Epping, NSW 1710, Australia}
\affiliation{Department of Computer Science, Math, Physics, \& Statistics, University of British Columbia, Kelowna, BC V1V 1V7, Canada}

\author{Z. F. Tang}
\affiliation{Purple Mountain Observatory, Chinese Academy of Sciences, Nanjing 210023, China}
\affiliation{School of Astronomy and Space Sciences, University of Science and Technology of China, Hefei 230026, China}

\author{S. Johnston}
\affiliation{CSIRO Space and Astronomy, Epping, NSW 1710, Australia}

\author{G. Hobbs}
\affiliation{CSIRO Space and Astronomy, Epping, NSW 1710, Australia}

\author{R. Manchester}
\affiliation{CSIRO Space and Astronomy, Epping, NSW 1710, Australia}

\author{X. F. Wu}
\affiliation{Purple Mountain Observatory, Chinese Academy of Sciences, Nanjing 210023, China}
\affiliation{School of Astronomy and Space Sciences, University of Science and Technology of China, Hefei 230026, China}

\author{P. Jiang}
\affiliation{National Astronomical Observatories, Chinese Academy of Sciences, Beijing 100101, China}
\affiliation{CAS Key laboratory of FAST, National Astronomical Observatories, Chinese Academy of Sciences, Beijing 100101, China}

\author{Y. F. Huang}
\affiliation{School of Astronomy and Space Science, Nanjing University, Nanjing 210023, China}

\author{Y. C. Zou}
\affiliation{School of Physics, Huazhong University of Science and Technology, Wuhan, 430074, China}

\author{Z. G. Dai}
\affiliation{Department of Astronomy, University of Science and Technology of China, Hefei 230026, China}

\author{B. Zhang}
\affiliation{Department of Physics and Astronomy, University of Nevada, Las Vegas, NV 89154, USA}

\author{D. Li}
\affiliation{National Astronomical Observatories, Chinese Academy of Sciences, Beijing 100101, China}
\affiliation{Research Center for Intelligent Computing Platforms, Zhejiang Laboratory, Hangzhou 311100, China}
\affiliation{University of Chinese Academy of Sciences, Beijing 100049, China}

\author{Y. P. Yang}
\affiliation{South-Western Institute for Astronomy Research, Yunnan University, Kunming 650504, China}
\affiliation{Purple Mountain Observatory, Chinese Academy of Sciences, Nanjing 210023, China}

\author{S. Dai}
\affiliation{School of Science, Western Sydney University, Locked Bag 1797, Penrith NSW 2751, Australia}

\author{C.M. Chang}
\affiliation{Purple Mountain Observatory, Chinese Academy of Sciences, Nanjing 210023, China}
\affiliation{School of Astronomy and Space Sciences, University of Science and Technology of China, Hefei 230026, China}

\author{Z. C. Pan}
\affiliation{National Astronomical Observatories, Chinese Academy of Sciences, Beijing 100101, China}
\affiliation{CAS Key laboratory of FAST, National Astronomical Observatories, Chinese Academy of Sciences, Beijing 100101, China}

\author{J. G. Lu}
\affiliation{National Astronomical Observatories, Chinese Academy of Sciences, Beijing 100101, China}
\affiliation{CAS Key laboratory of FAST, National Astronomical Observatories, Chinese Academy of Sciences, Beijing 100101, China}

\author{J. J. Wei}
\affiliation{Purple Mountain Observatory, Chinese Academy of Sciences, Nanjing 210023, China}

\author{Y. Li}
\affiliation{Purple Mountain Observatory, Chinese Academy of Sciences, Nanjing 210023, China}

\author{Q. W. Wu}
\affiliation{School of Physics, Huazhong University of Science and Technology, Wuhan, 430074, China}

\author{L. Qian}
\affiliation{National Astronomical Observatories, Chinese Academy of Sciences, Beijing 100101, China}
\affiliation{CAS Key laboratory of FAST, National Astronomical Observatories, Chinese Academy of Sciences, Beijing 100101, China}

\author{P. Wang}
\affiliation{National Astronomical Observatories, Chinese Academy of Sciences, Beijing 100101, China}
\affiliation{Institute for Frontiers in Astronomy and Astrophysics, Beijing Normal University, Beijing 102206, China}

\author{S. Q. Wang}
\affiliation{Xinjiang Astronomical Observatory, Chinese Academy of Sciences, Science 1-Street, Urumqi, Xinjiang 830011, China }

\author{Y. Feng}
\affiliation{Research Center for Intelligent Computing Platforms, Zhejiang Laboratory, Hangzhou 311100, China}

\author{L. Staveley-Smith}
\affiliation{International Centre for Radio Astronomy Research, University of Western Australia, Crawley, WA 6009, Australia}
\affiliation{ARC Centre of Excellence for All Sky Astrophysics in 3 Dimensions (ASTRO 3D)}



\begin{abstract}
Rotating Radio Transients (RRATs) are neutron stars that emit sporadic radio bursts. We detected 1955 single pulses from RRAT~J1913$+$1330 using the 19-beam receiver of the Five-hundred-meter Aperture Spherical Radio Telescope (FAST). These pulses were detected in 19 distinct clusters, with 49.4\% of them occurring with a waiting time of one rotation period.
The energy distribution of these individual pulses exhibited a wide range, spanning three orders of magnitude, reminiscent of repeating fast radio bursts (FRBs).
Furthermore, we observed abrupt variations in pulse profile, width, peak flux, and fluence between adjacent sequential pulses. 
These findings suggest that this RRAT could be interpreted as a pulsar with extreme pulse-to-pulse modulation. The presence of sequential pulse trains during active phases, along with significant pulse variations in profile, fluence, flux, and width, should be intrinsic to a subset of RRATs.  
Our results indicate that J1913+1330 represents a peculiar source that shares certain properties with populations of nulling pulsars, giant pulses, and FRBs from different perspectives. The dramatic pulse-to-pulse variation observed in J1913+1330 could be attributed to unstable pair creation above the polar cap region and the variation of the site where streaming pairs emit coherently.
Exploring a larger sample of RRATs exhibiting similar properties to J1913+1330 has the potential to significantly advance our understanding of pulsars, RRATs, and FRBs.
\end{abstract}

\keywords{radio transient --- RRATs --- individual pulses}


\section{Introduction} \label{sec:intro}
Since the discovery of 11 Rotating Radio Transients (RRATs) in 2006~\citep{McLaughlin06}, only around 100 such events are known until now~\footnote{See \url{http://astro.phys.wvu.edu/rratalog.}}. 
RRATs were identified by their sporadic radio bursts. Unlike the repeating Fast Radio Bursts(FRBs), which have been localized to extragalactic sources, RRATs are believed to originate from within the Milky Way based on their dispersion measures (DMs). By assuming the existence of underlying periodicities, periods can be derived by calculating the greatest common divisor of the time intervals between burst arrivals. Most RRATs with measured periods have periods of around one second, which aligns with typical rotation periods of neutron stars. 

Notably, the RRAT~J1819$-$1458 has been associated with a soft X-ray counterpart exhibiting thermal emission at $\sim$140\,eV~\citep{Reynolds06}. This finding suggested a cooling neutron star as the central engine of the RRAT. Furthermore, the high inferred brightness temperatures and plausible period increases~\citep{McLaughlin09} have led to the general belief that RRATs are a special type of pulsar. It is worth noting that some regular pulsars affected by intense radio frequency interference (RFI) or severe scintillation could be identified as RRATs if the initial observations do not have a sufficiently long search duration~\citep{Burke-Spolaor10}.   

However, the sporadic emission behaviour of the majority of RRATs requires new theories to explain. While some researchers have proposed that RRATs constitute a new population of pulsars~\citep{McLaughlin06, Keane08, McLaughlin09}, RRATs can be more readily interpreted as extreme nulling pulsars that are only ``on'' for less than a pulse period~\citep{Burke-Spolaor10, Burke-Spolaor13}, or as pulsars with extreme pulse-to-pulse modulation, with their normal emissions falling below the sensitivity thresholds of current telescopes~\citep{Weltevrede06}.  

RRAT~J1913$+$1330, originally known as J1913$+$1333, is one of the initial RRAT discoveries~\citep{McLaughlin06}. 
The period ($P_0$) and spin-down properties of J1913+1330 resemble those of typical radio pulsars, with $P_0 \approx 0.923$\,s, a surface magnetic field of 2.8$\times 10^{12}$\,G, and a spin-down energy loss rate of 4.2$\times 10^{32}$\,erg\,s$^{-1}$~\citep{McLaughlin09,Bhattacharyya18}.
Over the decade following its discovery, only sporadic individual pulses were detected, with an event rate of only a few per hour~\citep{McLaughlin06, McLaughlin09, Palliyaguru11, Shapiro-Albert18, Caleb19}, which is a typical RRAT behaviour. 
However, recent observations have revealed a ``weak persistent emission mode'' for this source~\citep{Bhattacharyya18, Lu19}. The average pulse rate of the weak persistent mode is $\sim 64\,{\rm hr}^{-1}$, with an average flux roughly 50 times fainter than the strong RRAT pulses.
This behaviour has never been found in other RRATs~\citep{Bhattacharyya18}, but resembles that of certain pulsars known to undergo mode change~\citep{Young15} or exhibit giant pulses~\citep{Kuzmin07}. 
Moreover, due to insufficient observation to distinguish individual pulses within the so-called ``weak persistent emission mode'', it remains unclear whether this mode is indeed the same as the strong RRAT pulses but with significantly lower intensities.

In this paper, we present the detection and analysis of numerous individual pulses from J1913$+$1330 using the Five-hundred-meter Aperture Spherical radio Telescope (FAST)~\citep{Jiang19}.
The observation details and data analysis are described in Section~\ref{sec:obs_data}. Section~\ref{sec:res} presents the properties of the detected pulses, while Section~\ref{sec:dis} presents a discussion of our findings. Finally, we conclude in Section~\ref{sec:con}.

\section{Observation and Data Analysis} \label{sec:obs_data}
To investigate the detailed emission properties of J1913$+$1330, we conducted a series of five observations using the FAST telescope~\citep{Jiang19} on August 17, December 16 and 25 of 2019, September 27 of 2021, and January 14 of 2022, totalling 8.9 hr.
We used the 19-beam receiver, which covers a frequency range of 1000$-$1500\,MHz with 4096 channels. 
The channelized signals were recorded using the Reconfigurable Open Architecture Computing Hardware generation 2~\citep[ROACH 2,][]{Hickish16} and sampled at 8-bit, stored in the PSRFITS search mode format~\citep{Hotan04}. The sample time was set to 49.153$\mu$s. 
During the observations, on August 17, we observed J1913$+$1330 for 0.9 hr by the central beam of the 19-beam receiver with two polarizations and recorded data from all beams.
For the remaining observations, signals were recorded only from the central beam of the multibeam receiver, but with four polarizations. 
The observation durations for December 16 and 25 of 2019, September 27 of 2021, and January 14 of 2022 were 3.0, 1.5, 1.6, and 1.9 hr, respectively. 
%calibration
Prior to each observation, a 10K equivalent noise-switched calibration signal was recorded to enable calibration of the results.

The data collected from the FAST radio telescope were processed using two individual search pipelines based on two pulsar/FRB single pulse searching packages: \emph{\sc presto}\footnote{\url{http://www.cv.nrao.edu/~sransom/presto/}}~\citep{Ransom01} and \emph{\sc heimdall}\footnote{\url{https://sourceforge.net/projects/heimdall-astro/}}. In both pipelines, the data sets were dedispersed in a range of DM values from 165 to 185\,cm$^{-3}\,$pc, with a step size of $0.01$\,cm$^{-3}$. Single pulse candidates with a signal-to-noise ratio (S/N) greater than seven were recorded and visually inspected. A candidate was identified as a detected pulse if it had an S/N above seven in {\sc presto} or {\sc heimdall} and showed a suggestive sweep in the de-dispersed frequency-time plane.

To scale our data to $T_{\rm sys}$ units, we utilized the 10K equivalent noise switched calibration signal prior to each observation.
We calibrated the flux density ($S$) of single pulses from Kelvin units to mJy by considering the zenith angle-dependent gain of the FAST telescope. This gain can be estimated using a function derived from observations of stable radio sources~\citep{Jiang20}. 
The standard deviation of the applied gains consistently remains around 3\% for all detections, with a maximum variation of approximately 14\%.
For the majority of clusters, the pulses were calibrated using gains of around $16 {\rm K/Jy}$. This is because nearly 90\% of our pulses were detected at zenith angles below 26.4 degrees, corresponding to a stable gain.
To compute the fluence ($F$), we integrated the pulse flux above the baseline, while the effective width ($W_{\rm ef}$) was determined by dividing the fluence by the pulse peak flux ($S_{\rm peak}$).

Our polarization data collected on MJD 58833, 58842, 59484 and 59593 were calibrated using the {\sc psrchive} software package~\citep{Straten2012}. The differential gain and phase between the receivers were corrected using the noise diode signal injected prior to each observation. 
We measure the rotation measures (RMs) for pulses with S/N $\geqslant 20$ using the {\sc rmfit} program to search for a peak in the linearly polarized flux ($L = \sqrt{Q^2 + U^2}$), in the range of RM values from $-1000$ to 1000\,rad\,m$^{-2}$, with a step size of $0.1$\,rad\,m$^{-2}$.

\section{Results}\label{sec:res}

\subsection{Single pulses detection}
\begin{figure*}
  \centering
  \includegraphics[width=160mm]{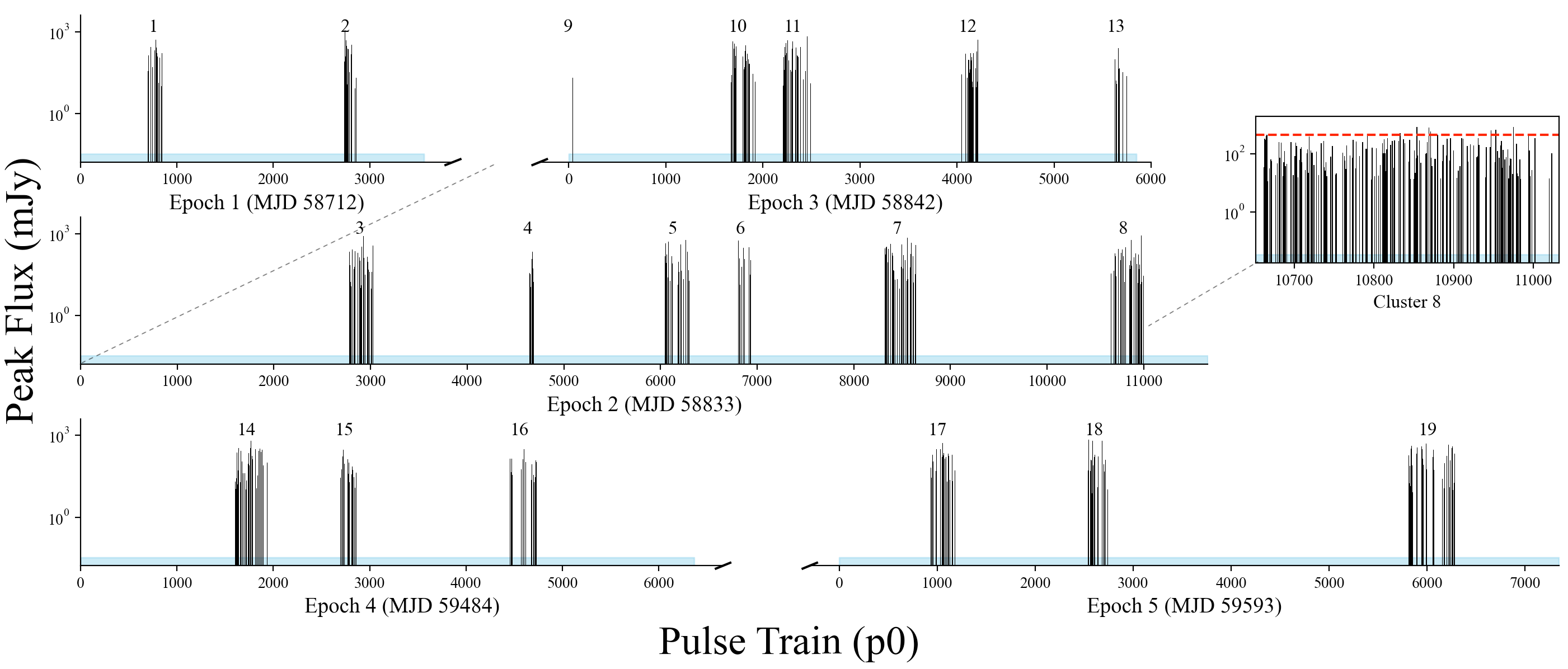}
  \caption{{\bf Peak flux of 1955 single pulses of RRAT~J1913+1330 in 19 clusters from 5 observation epochs.} The X-axis represents the pulse train number relative to the start of the observation, measured in units of the rotating period ($P_0$). Each observation epoch is indicated by a blue background, representing a flux level of 0.034\,mJy. This flux level corresponds to the peak flux limit during the nulling phase, and it is 1/194 of the weakest detected single pulse. The cluster number is indicated above each pulse. The red dashed line in the inset of Cluster 8 represents the upper limit for strong pulses, set at 460\,mJy.} 
 \label{segs}
\end{figure*}

During a total of 8.9 hr of observations, we detected 1955 individual pulses, corresponding to a pulse rate of $219\,{\rm hr}^{-1}$. 
Hereinafter we refer to the observation segment containing pulse clusters as the active phase. 
The pulses detected during the active phases formed 19 pulse clusters, as illustrated in Figure~\ref{segs}. 
These active phases lasted from $1.05$ to $7.62$ min (equivalent to $68-495$ $P_0$), with nulling phases between them ranging from $4.36$ to $46.85$ min (equivalent to $283-3044$ $P_0$).
Cluster 9 was detected $\sim$ 37\,s after the start of the observation, so our observation may only cover the end of this active phase, resulting in the detection of only one burst in this cluster.
For other clusters, whose active phases were fully covered, the number of pulses varied from 23 to 211.

Among the detected pulses, 49.4\% were with a waiting time of $1 P_0$. We estimated the pulse-on fraction of the active phase to be $0.42\pm0.06$. 
The effective width, peak flux and fluence of the single pulses ranged from 0.15 to 17.29\,ms, 0.006 to 1.84\,Jy and 0.004 to 5.13\,Jy\,ms, respectively. The brightest pulse reached a peak flux close to the previously observed maximum of 2.04\,Jy~\citep{McLaughlin09}. 
Out of the 1955 detected pulses, 103 pulses had peak fluxes above the reported average peak flux of 460\,mJy from J1913$+$1330~\citep{McLaughlin09}, and we will refer to these as ``strong pulse'' hereafter. 
The duration, number of detected pulses, number of strong pulses, minimum fluence ($F_{\rm min}$), maximum fluence ($F_{\rm max}$) and the maximum fluence contrast ($F_{\rm max}$/$F_{\rm min}$) are listed in Table~\ref{table:rate3}.

\begin{table*}
\caption{{\bf Properties of 19 pulse clusters of RRAT J1913+1330 detected by FAST during 5 epochs of observations.}}
\renewcommand\arraystretch{1.2}    %high
%\begin{center}
\begin{threeparttable}
\begin{tabular}{lcccccc}
\hline
\hline
Cluster/Epoch & Duration  &  Number of & Strong pulses &  $F_{\rm min}$ & $F_{\rm max}$ &  Maximum contrast \\
number & ($P_0$\tnote{a} ) &  pulses  & ($S \geq$ 460 mJy) & (Jy\,ms) & (Jy\,ms) &  $F_{\rm max}$/$F_{\rm min}$ \\
\hline
Cluster 1   &  228  & 80  & 7 &  0.0223 & 3.2548 & 145.8  \\%\hline
Cluster 2   &  123  & 61  & 6 &  0.0088 & 2.1179 & 240.5  \\
Epoch 1     &  351  & 141 & 13&  0.0088 & 3.2548 & 368.6  \\
\hline
Cluster 3   &  331  & 144 & 7 &  0.0143 & 5.1147 & 357.9  \\%\hline
Cluster 4   &  68   & 23  & 3 &  0.0159 & 3.4496 & 216.1  \\%\hline
Cluster 5   &  270  & 110 & 7 &  0.0154 & 2.6569 & 172.2  \\%\hline
Cluster 6   &  153  & 77  & 7 &  0.0164 & 2.8707 & 175.1  \\%\hline
Cluster 7   &  360  & 170 & 7 &  0.0046 & 4.0148 & 878.8  \\%\hline
Cluster 8   &  362  & 161 & 7 &  0.0274 & 2.5016 & 91.3   \\
Epoch 2     &  1544 & 685 & 38&  0.0046 & 5.1147 & 1119.9  \\
\hline
Cluster 9   &   -   & 1   &  - & 0.0836 & 0.0836 & -      \\%\hline
Cluster 10  &  254  & 103 & 8 &  0.0260 & 3.6239 & 139.5  \\%\hline
Cluster 11  &  313  & 138 & 4 &  0.0162 & 2.1370 & 132.1  \\%\hline
Cluster 12  &  232  & 108 & 8 &  0.0093 & 3.0279 & 326.2  \\%\hline
Cluster 13  &  123  & 45  & 4 &  0.0206 & 2.5151 & 122.2  \\
Epoch 3     &  922  & 395 & 24&  0.0093 & 3.6239 & 390.4  \\
\hline
Cluster 14  &  335  & 152 & 3 &  0.0039 & 1.4696 & 370.4  \\%\hline
Cluster 15  &  219  & 88  & 0 &  0.0056 & 1.4948 & 267.0  \\%\hline
Cluster 16  &  316  & 88  & 1 &  0.0111 & 2.4326 & 219.3  \\
Epoch 4     &  870  & 328 & 4 &  0.0039 & 2.4326 & 613.2 \\
\hline
Cluster 17  &  251  & 106 & 7 &  0.0131 & 2.4368 & 186.1  \\%\hline
Cluster 18  &  229  & 89  & 8 &  0.0135 & 3.3578 & 248.0  \\%\hline
Cluster 19  &  495  & 211 & 9 &  0.0076 & 3.5474 & 467.0  \\
Epoch 5     &  975  & 406 & 24&  0.0076 & 3.5474 & 467.0 \\
\hline
Full sample       &  4662 & 1955 & 103 & 0.0039 & 5.1147 & 1289.3 \\\hline 
\end{tabular}
  \begin{tablenotes}
        \footnotesize
        \item[a]The duration lengths presented here are given in units of the rotating period ($P_0$) of J1913+1330 during each observation.
     \end{tablenotes}
\end{threeparttable}
\label{table:rate3}
\end{table*}

\subsection{Nulling phase limit}

Between the active phases, there are quiescent nulling phases, as shown in Figure~\ref{segs}. 
We added the nulling phases of each observation to obtain the integrated pulse. However, no convincing candidate with an S/N greater than seven was detected. Limits on the phase-average flux density can be estimated as:
\begin{equation}
S_{\rm lim}=\frac{{\rm S/N}_{\rm min}\,T_{\rm sys}}{G \sqrt{{\Delta}{\nu}N_p{t}_{\rm obs}}}\sqrt{\frac{\delta}{1-\delta}},
\label{equ:limit}
\end{equation}
where $T_{\rm sys} \sim 20{\rm K}$ is the system temperature, $G \sim 16 {\rm K/Jy}$ is the telescope antenna gain, ${\Delta}{\nu} = 500$\,MHz is the observing bandwidth, $N_p =2$ is the sum of polarizations, $t_{\rm obs}$ is the integration time, $\delta$ is the pulse duty cycle (defined as the ratio of the pulse width to the period, and estimated by the pulse-on segments integrated profile to be 6\,ms/923\,ms $\sim$ 0.006), and ${\rm S/N}_{\rm min}$ is set to 7.0. 

For the longest observation in our dataset (Epoch 2), which encompassed a total nulling phase time of approximately 9343\,s, we obtained limits on the mean flux density, peak flux density, and fluence of approximately $0.22\,\mu$Jy, $33.86\,\mu$Jy and $2.0 \times 10^{-4}$ Jy\,ms, respectively. The peak flux and fluence are approximately 194 and 20 times fainter than the smallest values of the detected pulses, indicating the absence of pulses during the nulling phases.

\subsection{DM \& RM}
\begin{figure}
  \centering
  \includegraphics[width=80mm]{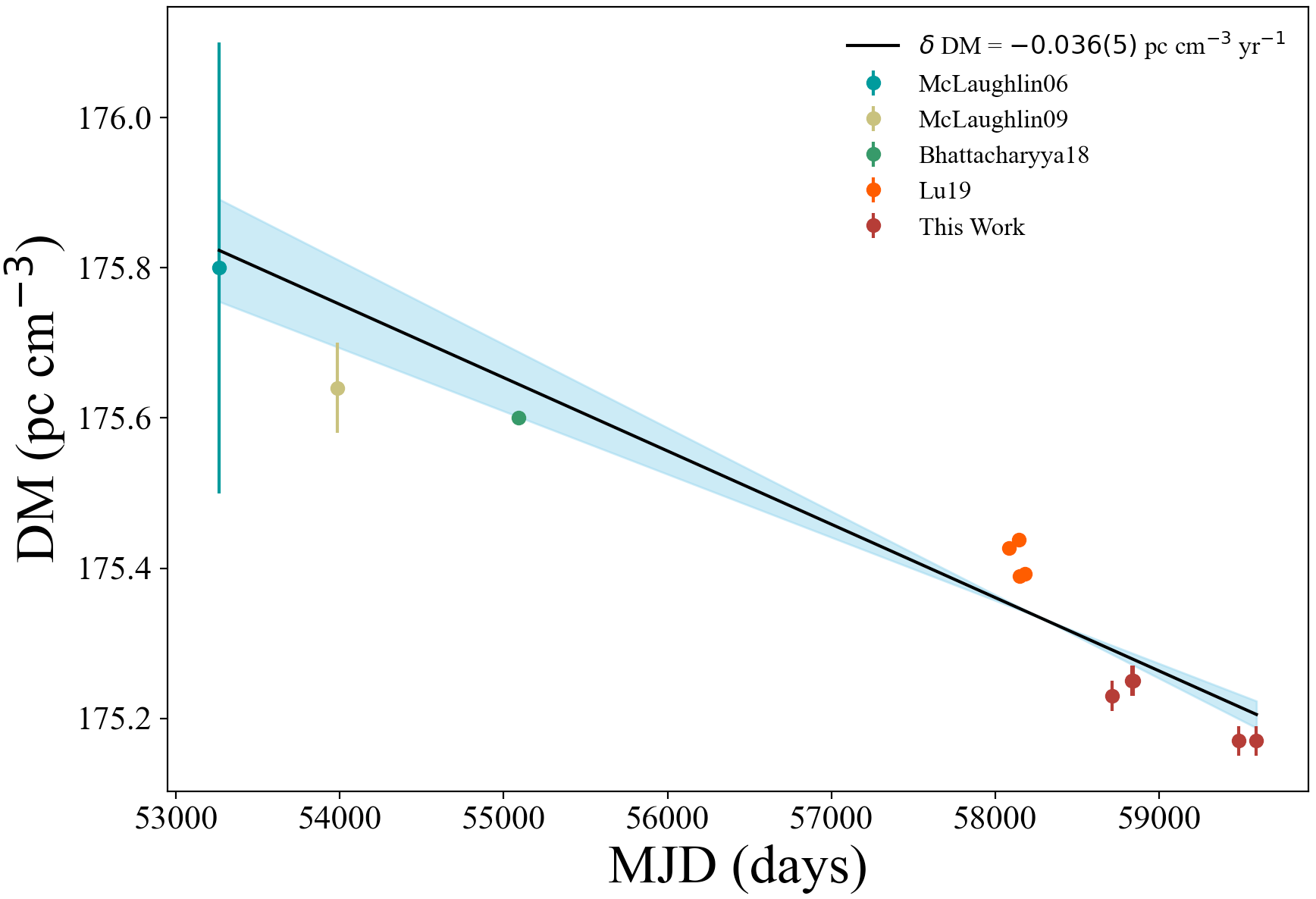}
  \caption{{\bf Comparison of DMs of J1913$+$1330 obtained in previous studies and this work.} The DM values of this work are smaller than those from the previous study, suggesting a possible evolution of $\delta {\rm DM} = -0.035\,{\rm pc\,cm}^{-3} {\rm yr}^{-1}$.  }
 \label{DM_decrease}
\end{figure}

The DM of each single pulse from J1913$+$1330 was derived by maximizing the S/N of its integrated pulse profile. Since there was no significant DM variation identified within each observation, we analysed the optimal DM value for all single pulses detected in each observation to obtain the best-fit value.  
Our analysis resulted in best-fit DM values of 175.23(2), 175.25(2), 175.25(2), 175.17(2) and 175.17(2) $\rm cm^{-3}\rm pc$ on MJD 58712, 58833, 58842, 59484 and 59593, respectively. 
The DM values for the last two observations were slightly smaller than those for the first three.
As shown in Figure~\ref{DM_decrease}, our DM values are smaller than those reported in previous studies~\citep{McLaughlin06,McLaughlin09,Bhattacharyya18,Lu19}, suggesting a possible decrease rate of $\delta {\rm DM} = -0.035 \pm 0.005\,{\rm pc\,cm}^{-3} {\rm yr}^{-1}$.

\begin{figure*}
\centering
\begin{tabular}{ccc}
\includegraphics[width=5.5cm]{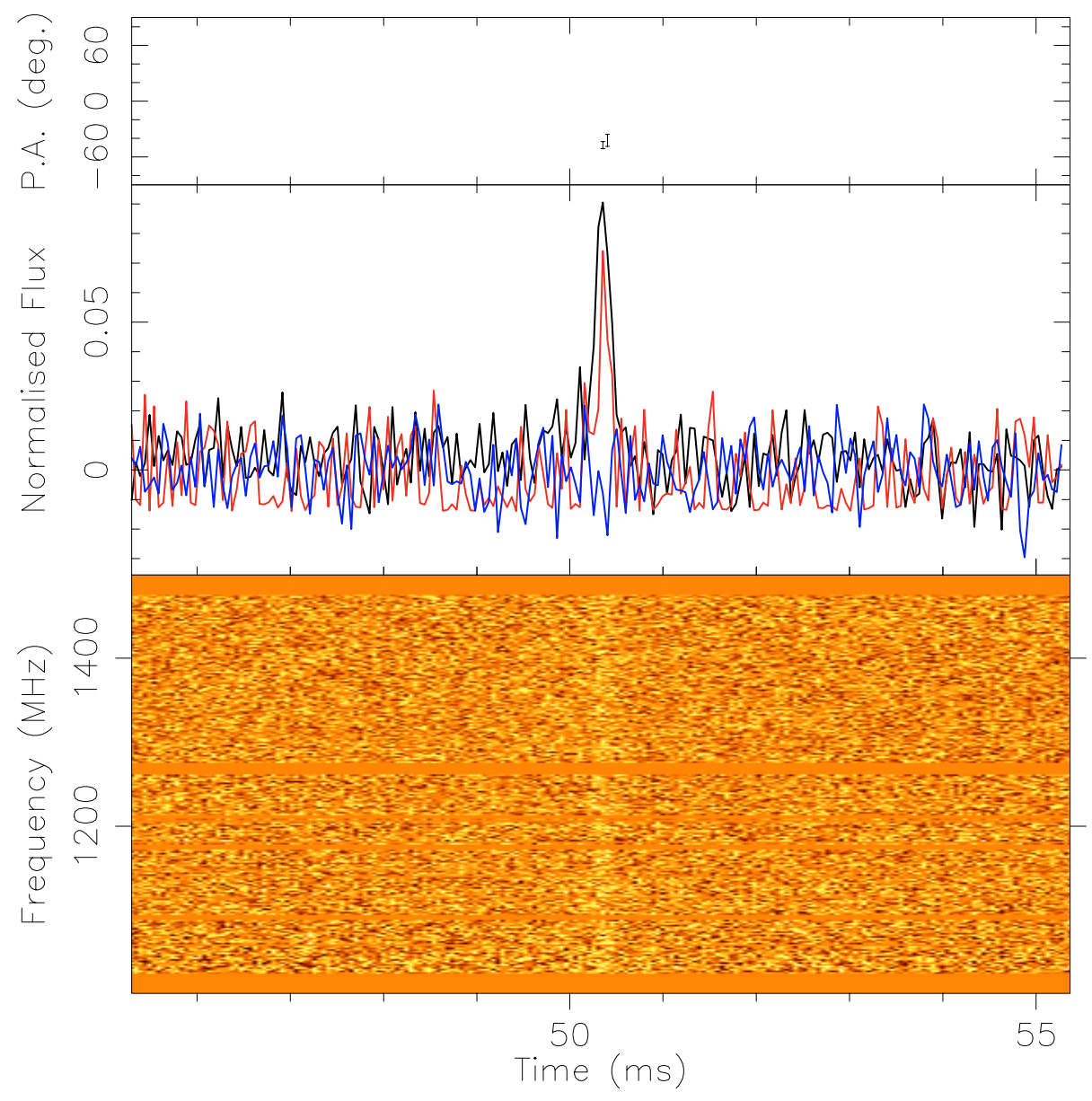} &
\includegraphics[width=5.5cm]{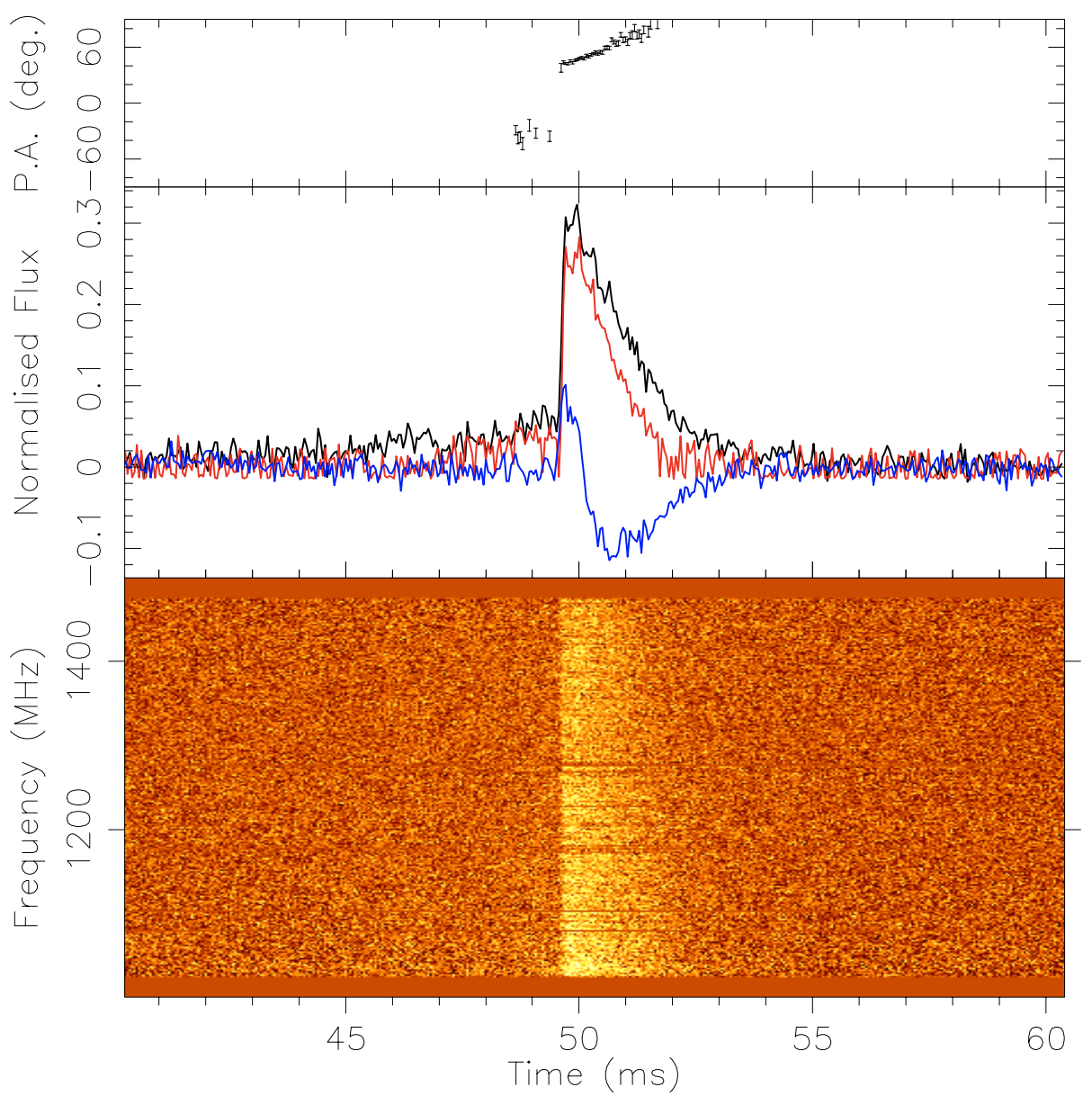} &  
\includegraphics[width=5.5cm]{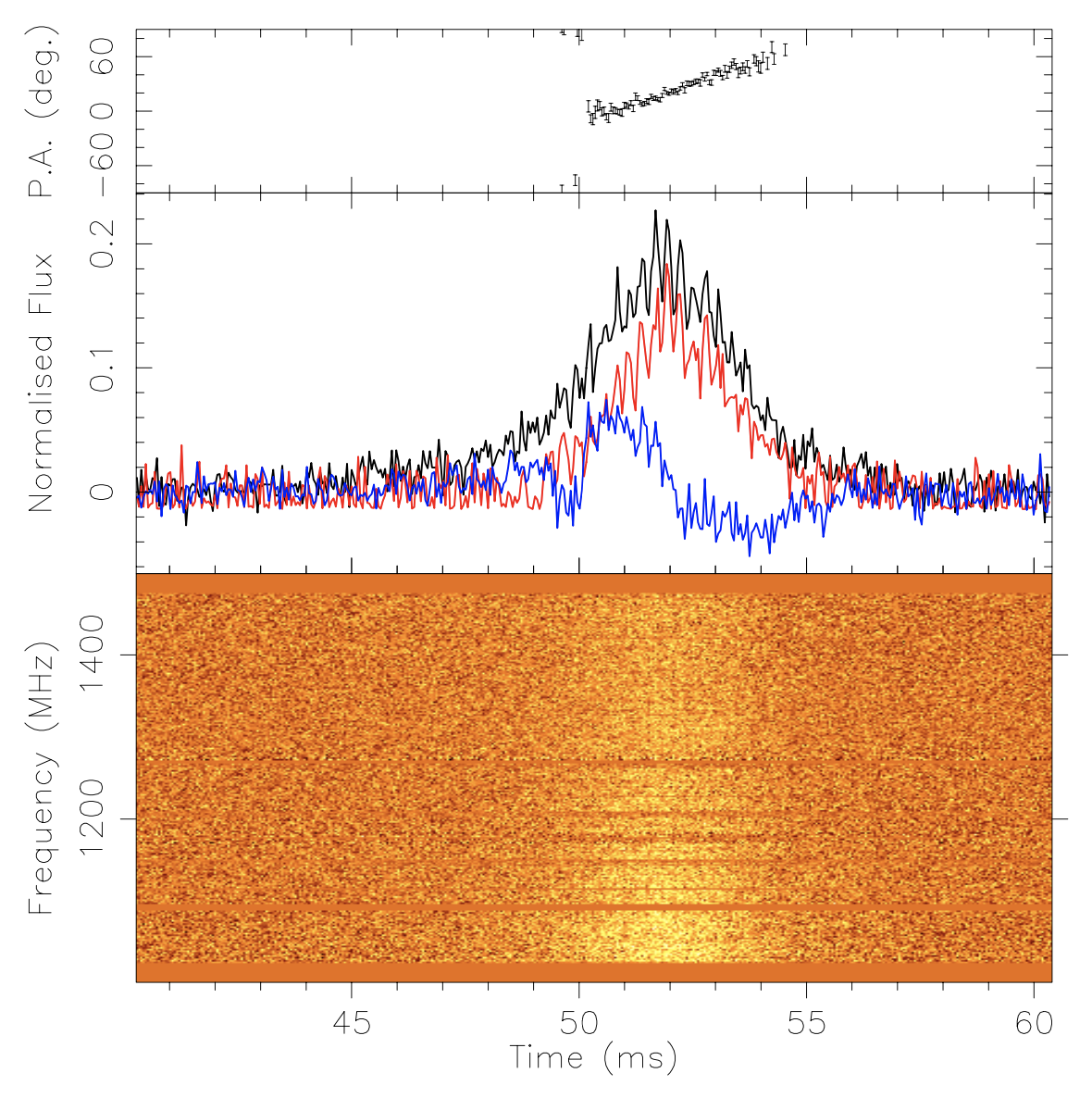}  \\
\includegraphics[width=5.5cm]{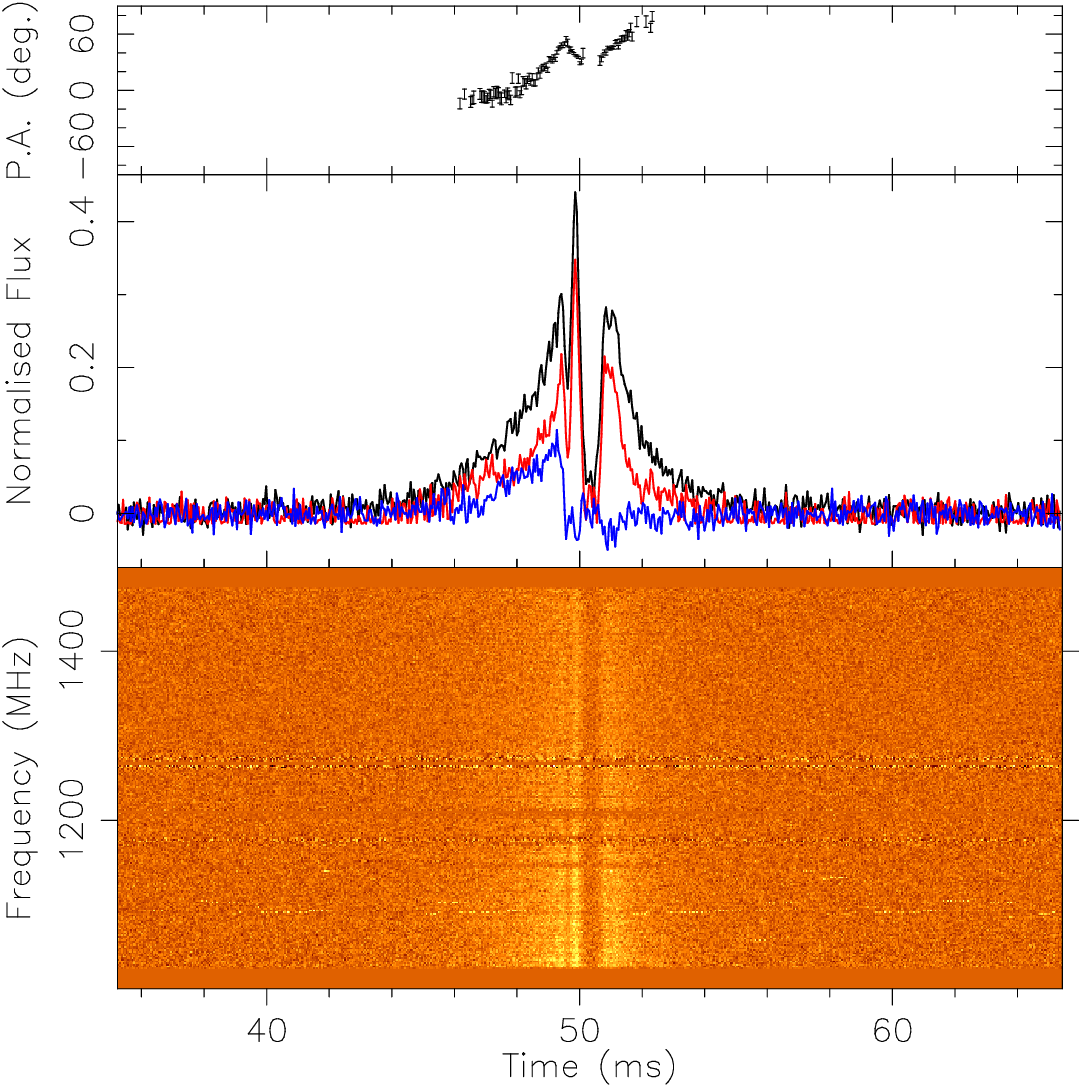} &
\includegraphics[width=5.5cm]{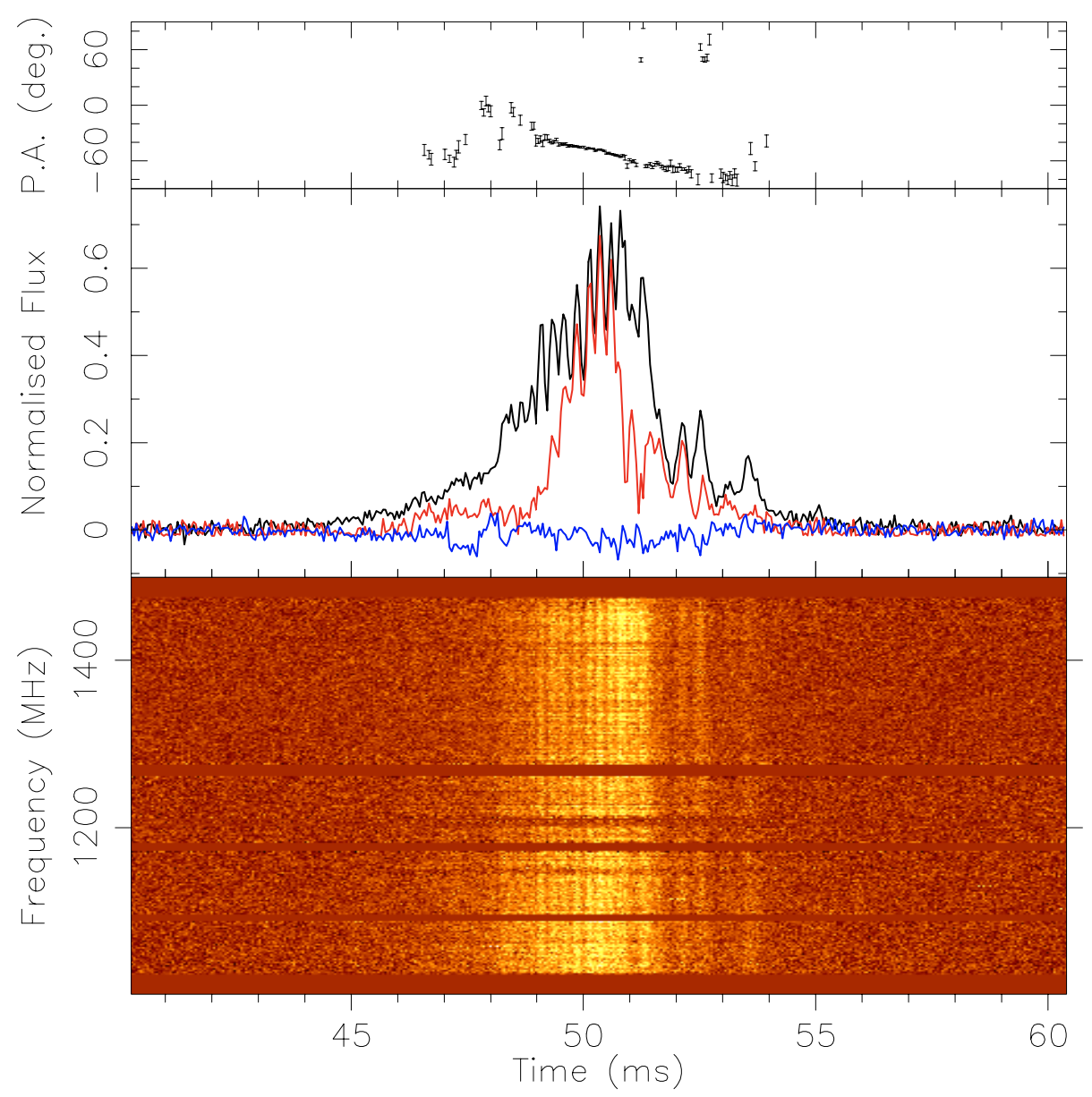} &
\includegraphics[width=5.5cm]{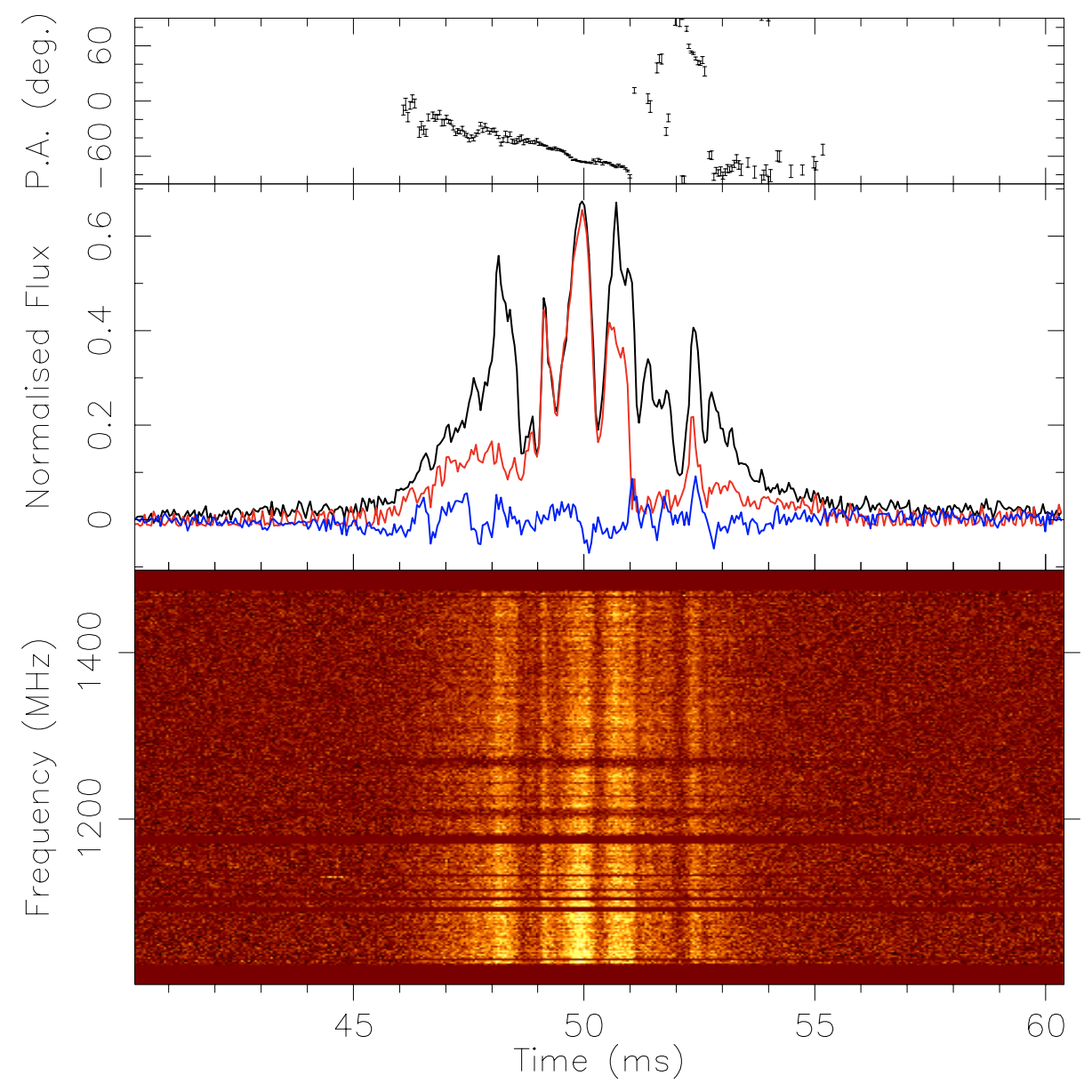}  \\
\includegraphics[width=5.5cm]{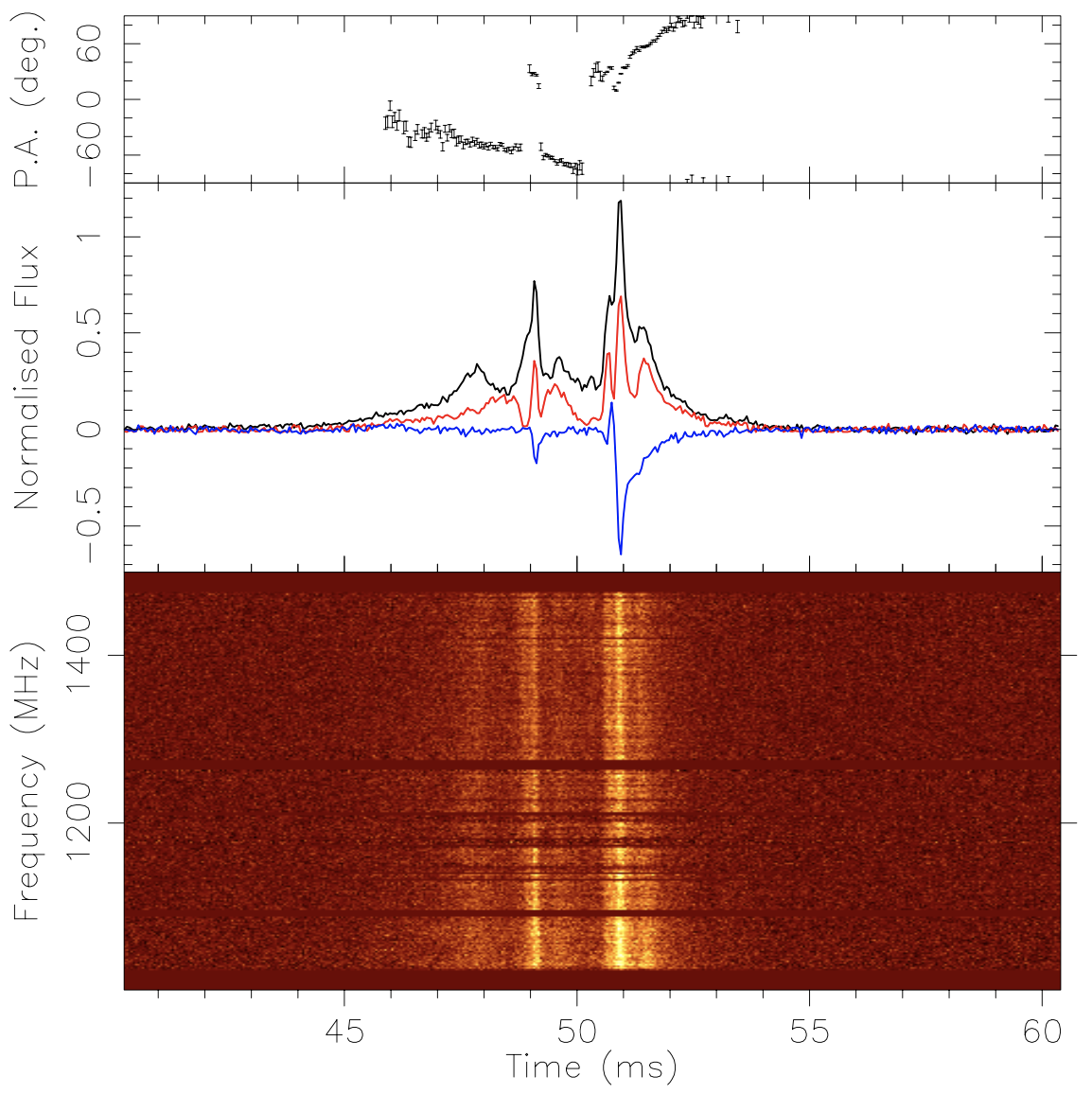} & 
\includegraphics[width=5.5cm]{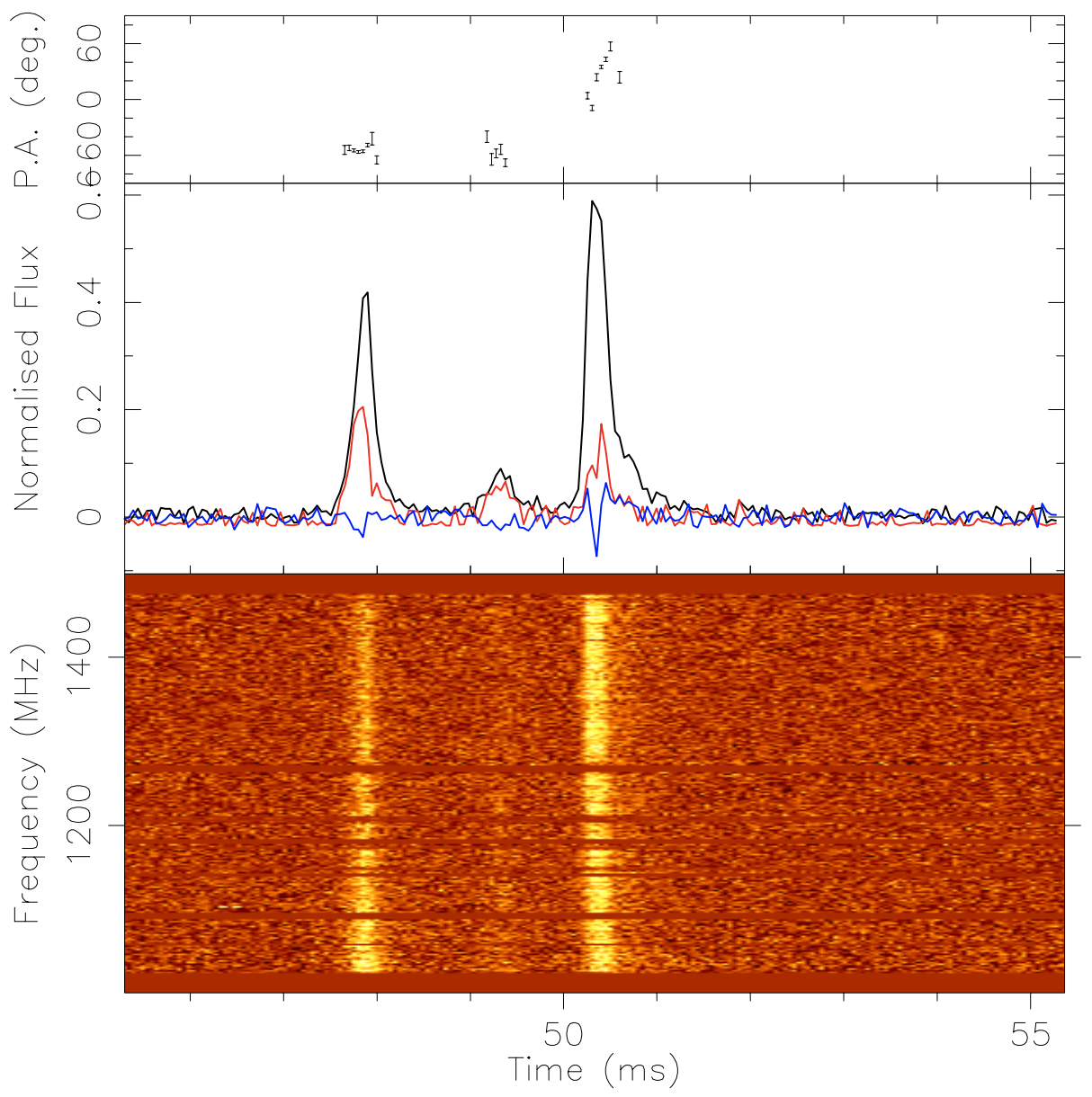} &  
\includegraphics[width=5.5cm]{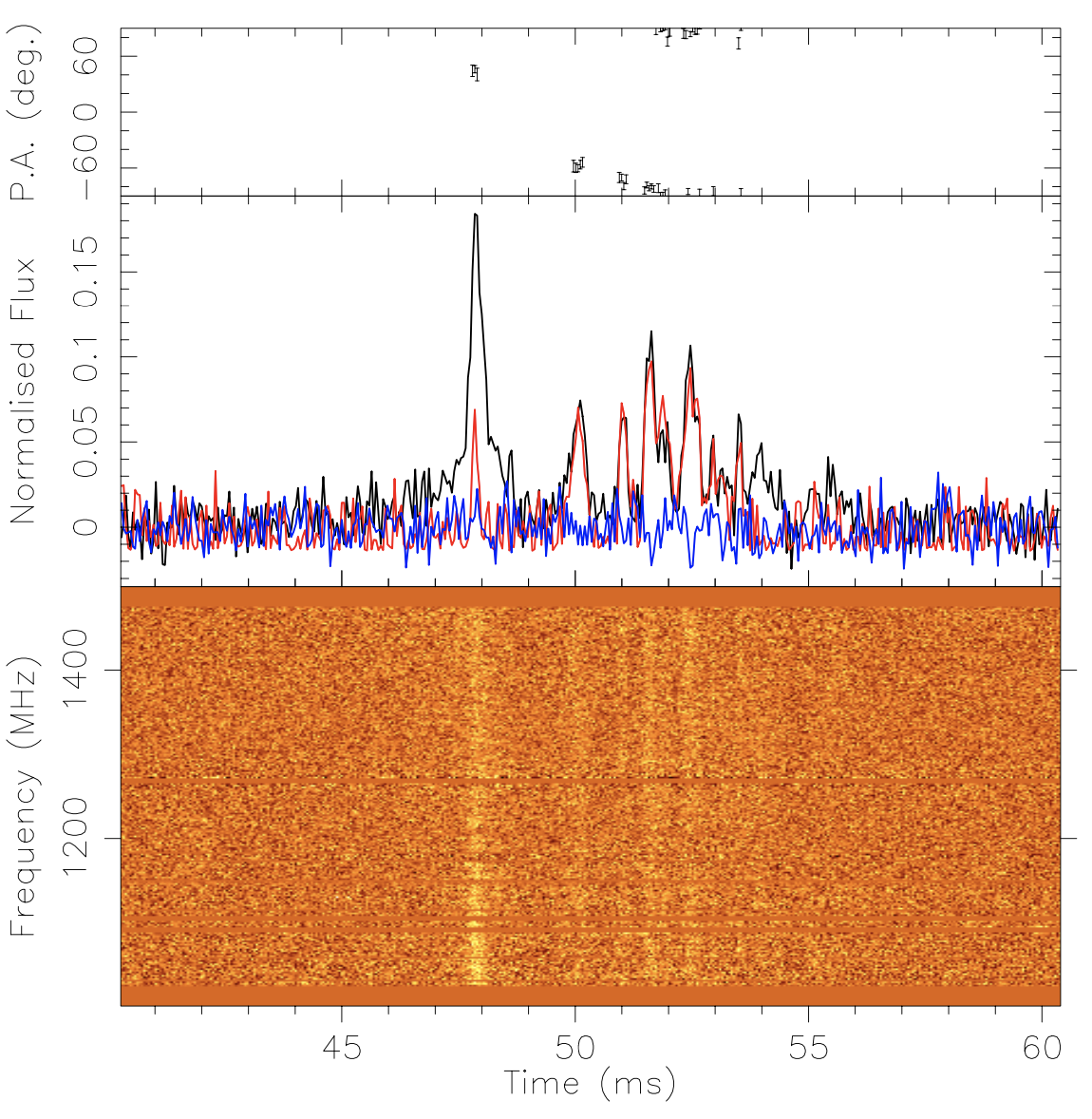} \\    
\end{tabular}
\caption{{\bf Polarization profiles of a sub-sample of single pulses from RRAT~J1913$+$1330.} Each subplot includes the following information: the upper panel displays the position angle of linear polarization at 1250\,MHz, the middle panel shows the polarization pulse profile, with black, red and blue curves representing total intensity, linear polarization, and circular polarization respectively, and the lower panel presents the dynamic spectra for the total intensity with a frequency resolution of 1.95\,MHz/channel and time resolution of 49.125\,$\mu$s/bin for all pulses.}
\label{f_micro}
\end{figure*}

The best estimate of RM for the detected bursts yielded a value of 932.8$\pm$6.6\,rad\,m$^{-2}$, slightly smaller than the previously reported value of 945$\pm$11\,rad\,m$^{-2}$~\citep{Caleb19}. 
All detected pulses in the four-polarization observations were RM corrected using the average RM value. 
Figure~\ref{f_micro} presents polarization profiles of a sample of single pulses from J1913$+$1330. 
The source exhibits a wide range of polarization angles (PA) and degrees of linear polarization, with a small fraction of pulses showing a high degree of circular polarization.
A detailed polarisation property analysis of all bursts will be presented in a companion paper.

\subsection{Single pulse profile \& Microstructure}
The profiles of single pulses from J1913$+$1330, as shown in Figure~\ref{f_micro}, exhibit significant variations.
During certain active periods, we detected only single-peak pulses with $W_{\rm ef}$ ranging from 0.15 to 3.63\,ms. 
The effective time resolution of our observations could be estimated by~\cite{Cordes03}: 
\begin{equation}
w_{\rm eff}=\sqrt{{w_{\rm DM}^2}+{w_{\Delta \rm DM}^2}+{w_{\rm sample}^2}+{w_{\rm scatt}^2}},
\label{equ:eff_wid}
\end{equation}
where $w_{\rm DM}$ represents the dispersion smearing ($\sim$ $53-177\,\mu$s at frequency of $1000-1500$\,MHz), $w_{\Delta \rm DM}$ accounts for the dedispersion error (negligible in our observation), $w_{\rm sample} = 49.153\,\mu$s is the sample time, and $w_{\rm scatt}$ represents the scattering smearing ($\sim 14-82\,\mu$s at frequency of $1000-1500$\,MHz~\footnote{The scattering smearing values were estimated using the empirical fit suggested in Equation (6) of~\cite{Cordes03}.}).
Our observation has an effective time resolution ranging from approximately 73 to 201\,$\mu$s for frequencies from 1.0 to 1.5\,GHz. 
It is worth noting that the width of the narrowest burst we detected is similar to the effective time resolution of our observation, indicating the possible existence of even narrower pulses from this source. 
More than 95\% of the detected pulses exhibit diverse microstructures, limited only by our effective time resolution.
In approximately 10\% of these pulses, multiple distinct individual pulses can be distinguished.
The largest effective width of the pulses we detected is 17.29\,ms.

\subsection{Times of arrival}

\begin{figure}
  \centering
 \includegraphics[width=80mm]{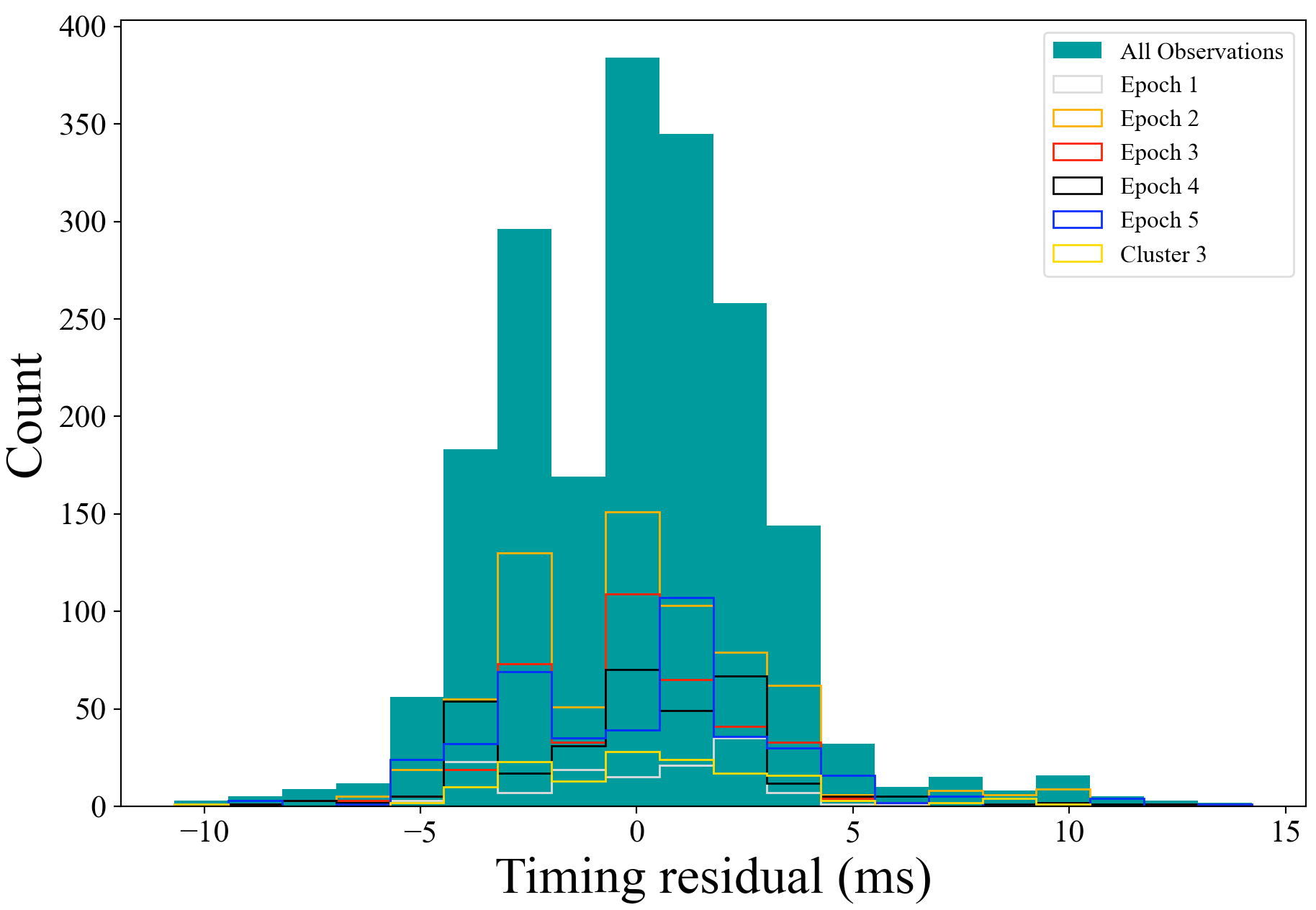}
  \caption{The timing residuals from the TOAs of the 1955 single pulses from RRAT~J1913$+$1330. The histogram represents the distribution of all observations, while for comparative purposes, we also display the distributions of the five Epochs and Cluster 3.}
\label{toa_residual}
\end{figure}

We measured the pulse times of arrival (TOAs) at the flux peak and then converted the local arrival times to barycentric arrival times using the TEMPO2 software package~\citep{Hobbs06}. The timing residuals of the TOAs for the 1955 single pulses showed a bimodal distribution, as shown in Figure~\ref{toa_residual}, with a maximum TOA differential of approximately 22.91\,ms observed. This differential is only slightly larger than our widest pulse of 17.29\,ms.

\begin{figure}
  \centering
 \includegraphics[width=80mm]{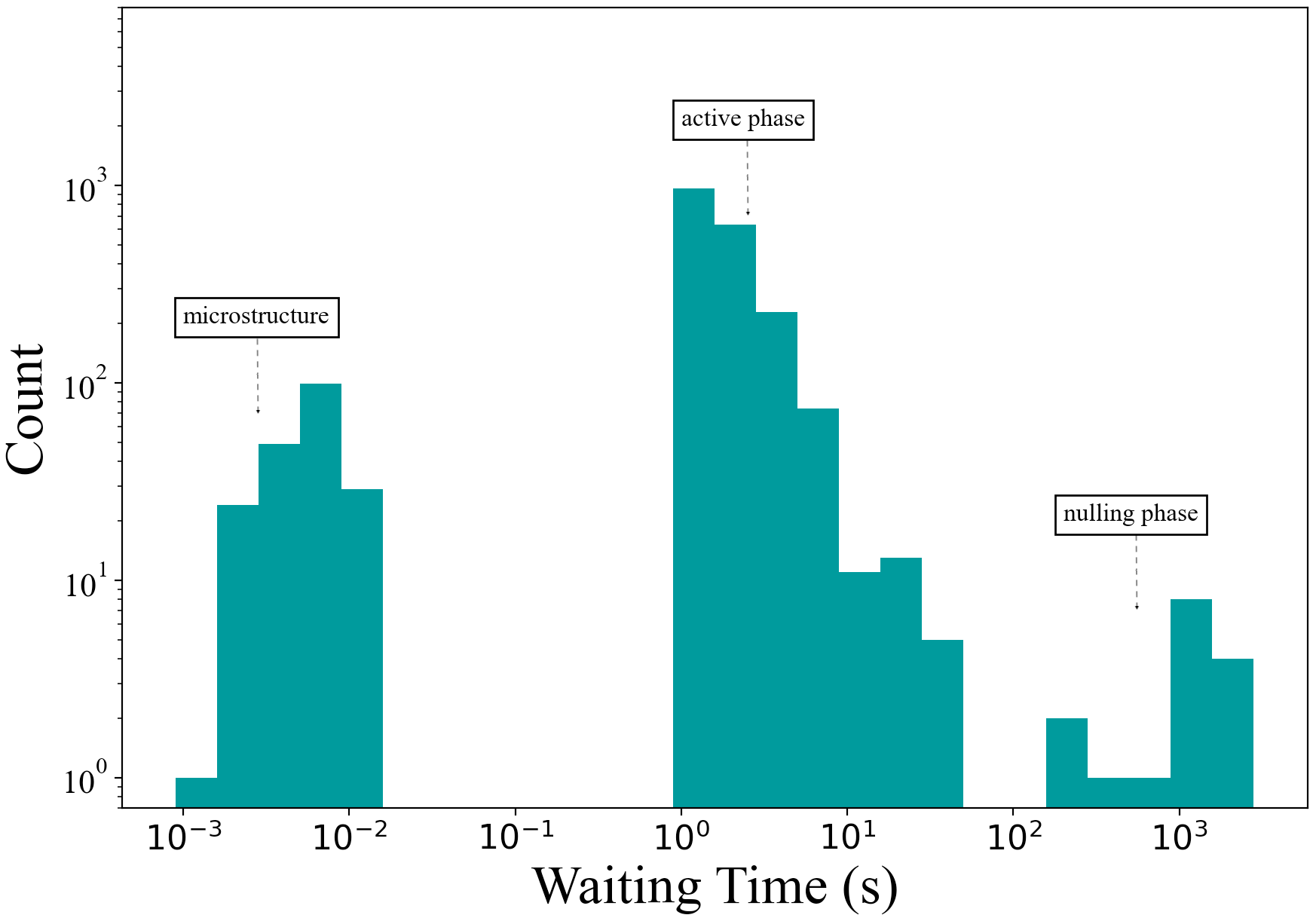}
  \caption{The waiting time distribution of RRAT~J1913$+$1330. The distinct individual pulses in one rotating period were regarded as different pulses in this analysis and a total of 2160 pulses were identified. 
  The distribution comprises three components: (1) timescale of $\sim$ 1000\,s representing the intervals between active phases, (2) timescale of $\sim$ 1\,s representing the waiting time within the active phases, and (3) timescale of $\sim$ 1\,ms representing the microstructure within one rotating period.  }
\label{wait_time}
\end{figure}

Of the 1955 single pulses, 49.4\% of them have waiting time $\sim 1P_0$.
Multiple pulses were observed in numerous active phases, and we extracted 2160 single pulses by distinguishing distinct individual pulses in the same rotational period as different pulses. The waiting time distribution of these 2160 pulses is presented in Figure~\ref{wait_time}. The distribution comprises three components: (1) timescale of $\sim$ 1000\,s representing the intervals between active phases, (2) timescale of $\sim$ 1\,s representing the waiting time within the active phases, and (3) timescale of $\sim$ 1\,ms representing the microstructure within one rotating period.

\section{Discussion} \label{sec:dis}
\subsection{Distribution of fluence, peak flux and width}

\begin{figure*}
\centering
\begin{tabular}{cc}
\includegraphics[width=81mm]{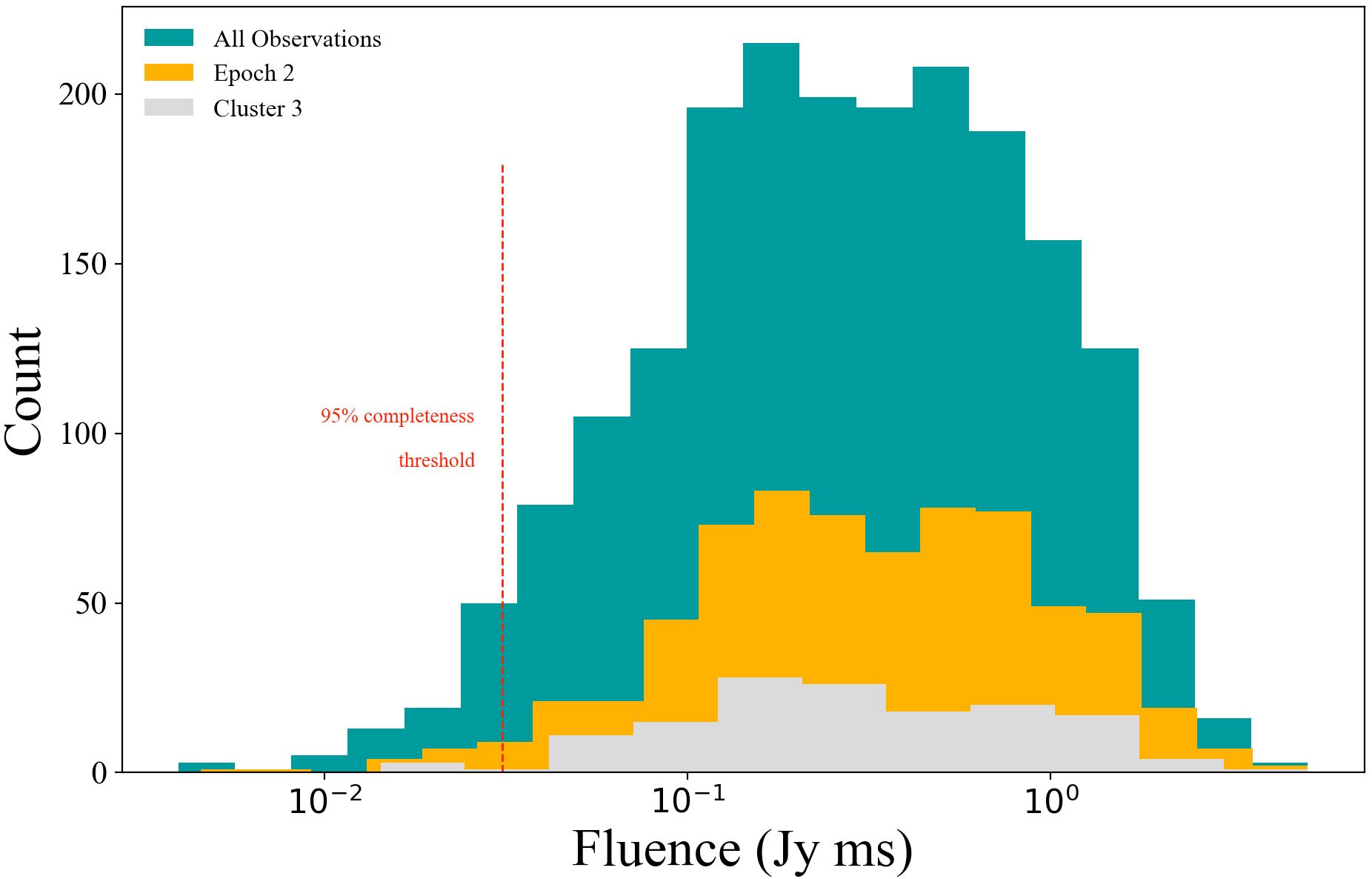}  &  
\includegraphics[width=80mm]{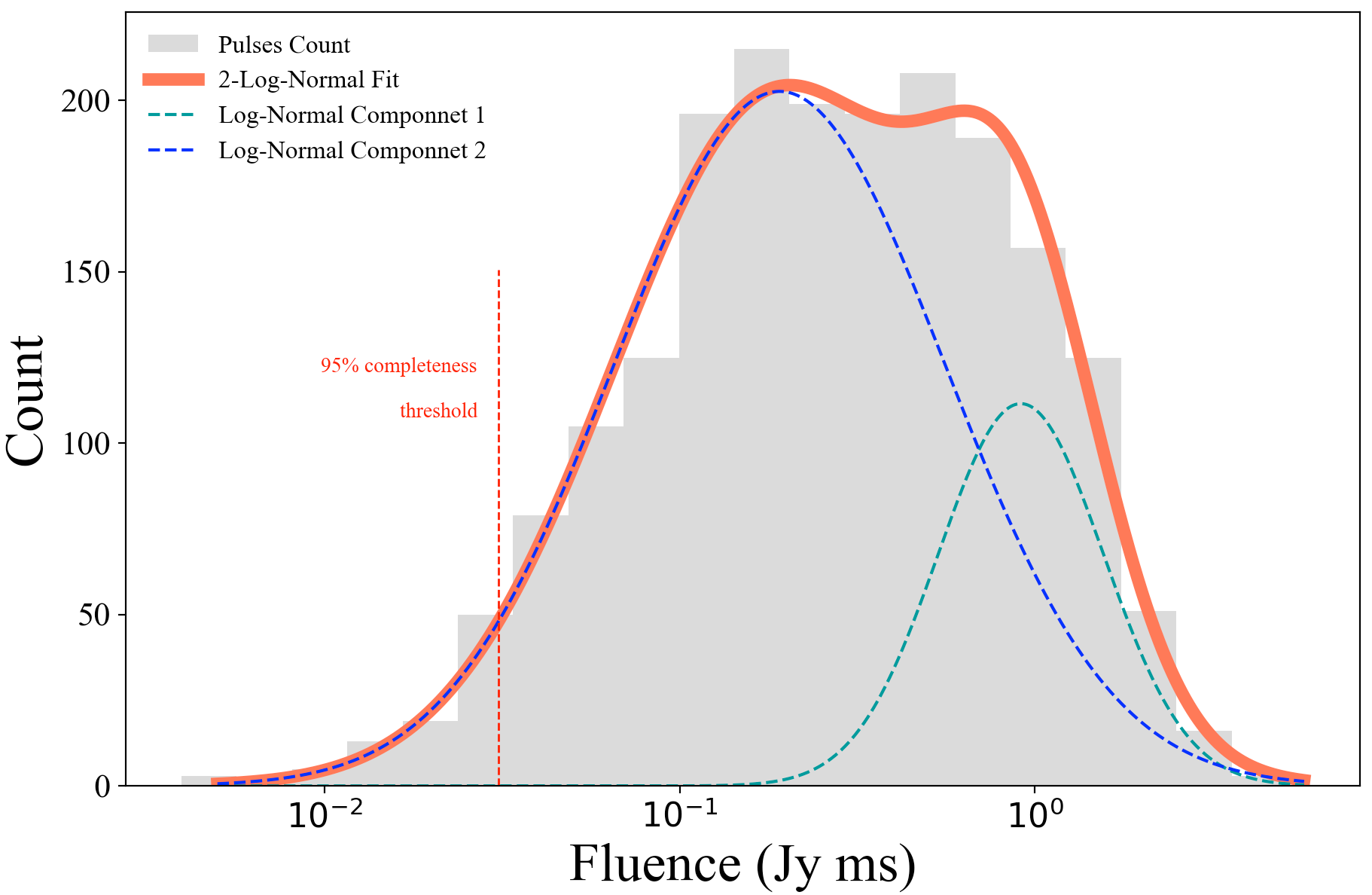} 
\end{tabular}
\caption{{\bf \emph{Left:} Specific fluence distribution of the single pulses from RRAT~J1913$+$1330.} The cyan histogram represents all 1955 pulses from our observations, while the orange histogram indicates the 685 pulses detected in Epoch 2, and the gray histogram represents the 144 pulses in Cluster 3. The 95 percent detection completeness threshold is shown by the red dashed line, corresponding to 0.031\,Jy\,ms for an assumed pulse width of 4\,ms.
{\bf \emph{Right:} Fluence distribution with the best two log-normal fit.} The histogram shows the distribution, and the red line represents the best fit using a model with two log-normal functions. The two log-normal distributions peak at 0.19 and 0.91\,Jy\,ms, respectively.}
\label{flue_dis}
\end{figure*}

The fluence distribution of the detected pulses from J1913$+$1330 is illustrated in Figure~\ref{flue_dis}. 
Notably, the fluence exhibits significant variations spanning over three orders of magnitude.
To account for the uncertainty introduced by flux calibration, we compared the fluence distribution of pulses in all observations with that of our longest observation (Epoch 2, lasting for approximately 3 hr) and Cluster 3 (lasting for approximately 5 min) within this observation.
It was expected that the variation in system temperature would be relatively small in these cases.
The distributions of a single epoch and a cluster were found to be consistent with the overall sample.
The maximum fluence contrast ($F_{\rm max}$/$F_{\rm min}$) of the detected pulses in all observations, Epoch 2 and Cluster 3 were 1289, 1120, and 358, respectively. 
It should be noted that similar variations in pulse fluence also exist in other individual clusters, as shown in Table \ref{table:rate3}. 
We utilized two log-normal functions to fit the fluence distribution of all the detected single pulses, with peaks at 0.19 and 0.91\,Jy\,ms. 
For convenience, we categorized pulses with fluence values below 0.19\,Jy\,ms as low energy, pulses with fluence values ranging from 0.19 to 0.91\,Jy\,ms as middle energy, and pulses with fluence values above 0.91\,Jy\,ms as high energy. 

The fluence distribution of J1913$+$1330 is noteworthy because broad energy distributions are atypical for regular pulsar single pulses~\citep{Burke-Spolaor12}, and intensity variations of more than one order were always attributed to the phenomenon of ``giant pulses''~\citep{Kuzmin07}.  
In our observations, even most low energy pulses are well above our 95\% detection completeness threshold of 0.031\,Jy\,ms\footnote{This threshold was determined based on simulated mock pulses search results~\cite{Li21}, taking into consideration an estimated mean pulse width of approximately 4 ms for our detections.}, indicating that we may have detected the majority of pulses from this source. Notably, there is no evidence supporting the presence of two distinct emissions, namely ``giant pulses'' and ``normal pulses''.

\begin{figure}
\centering
 \includegraphics[width=80mm]{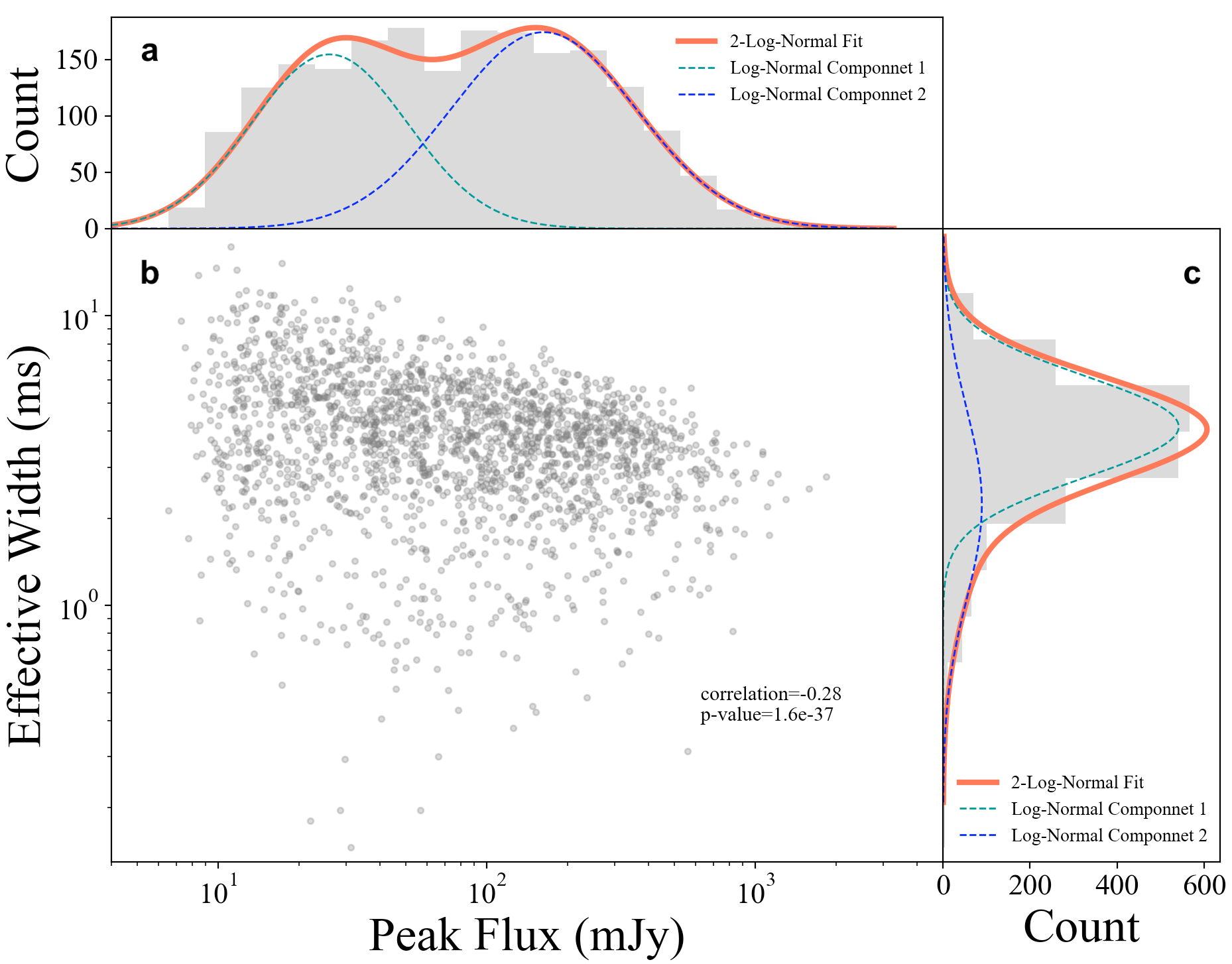} 
\caption{{\bf Effective width and peak flux density distribution for 1955 RRAT~J1913$+$1330 single pulses.} (a)Histogram showing the peak flux distribution. (b) Two-dimensional distribution of peak flux and pulse width, with the coefficient and p-value of the Spearman rank-order correlation presented. (c) Histogram displaying the effective width distribution. The histogram plots also include the best fits using two log-normal functions for the width and peak flux distributions.}
\label{f_wid_flx}
\end{figure}

\begin{figure*}
\centering
\begin{tabular}{cc}
\includegraphics[width=140mm]{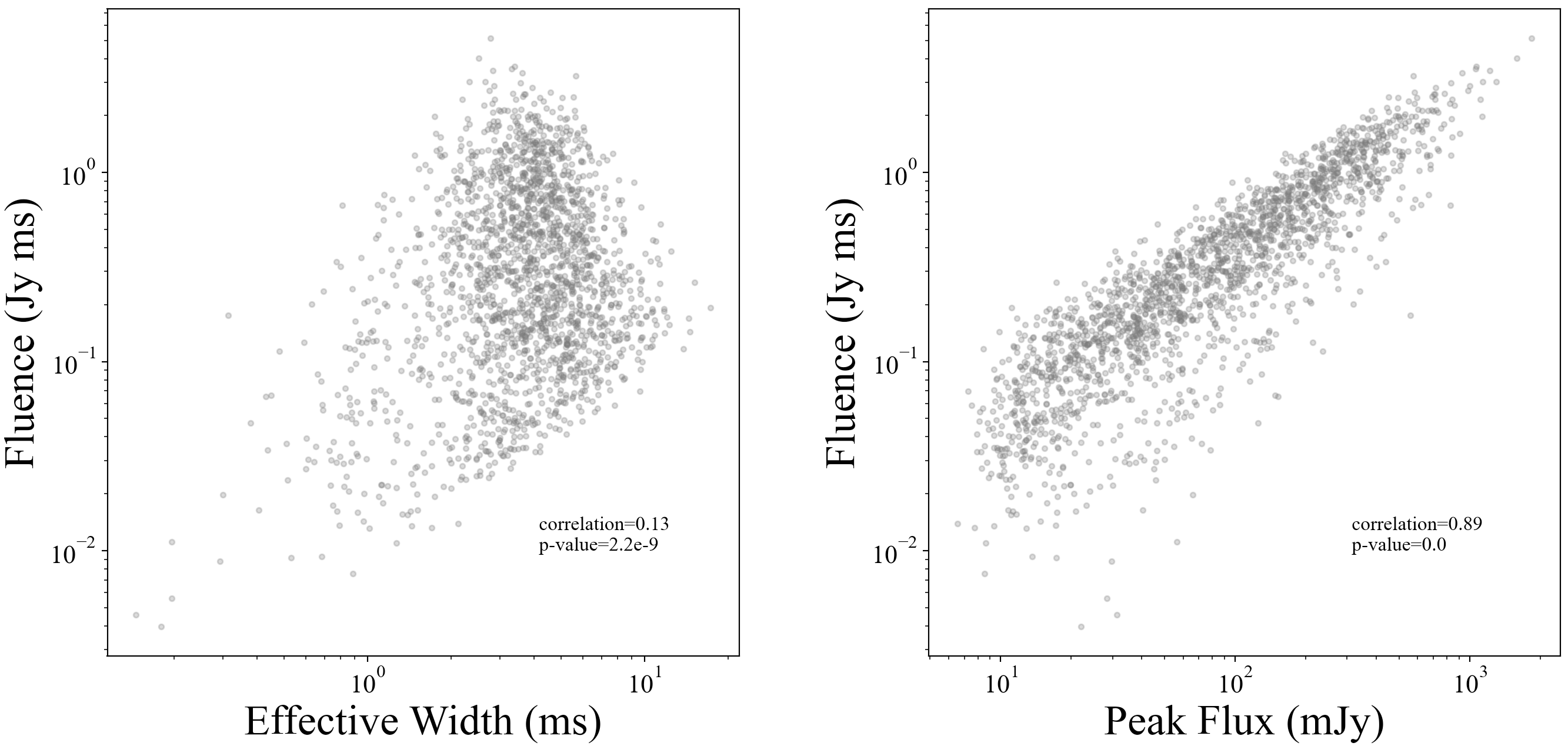}  
\end{tabular}
\caption{{\bf Scatter plots of effective width and fluence (\emph{Left}), and peak flux and fluence (\emph{Right}) for 1955 RRAT~J1913$+$1330 single pulses.} The coefficients and p-values of the Spearman rank-order correlation are presented in each plot.}
\label{f_wid_flx_flu}
\end{figure*}

The effective width and peak flux distributions of the 1955 single pulses of J1913$+$1330 are presented in Figure~\ref{f_wid_flx}.  
We also fitted these distributions using two log-normal functions, with maximum contrasts in flux and width (i.e., $S_{\rm max}/S_{\rm min}$ and $W_{\rm max}/W_{\rm min}$) being 281 and 118, respectively. 
As shown in Figure~\ref{f_wid_flx_flu}, a strong positive correlation between fluence and flux is observed, indicated by Spearman's correlation coefficient of 0.89, while the correlation between fluence and effective width is insignificant.

\subsection{Pulse-to-pulse variation}

\begin{figure}
  \centering
  \includegraphics[width=80mm]{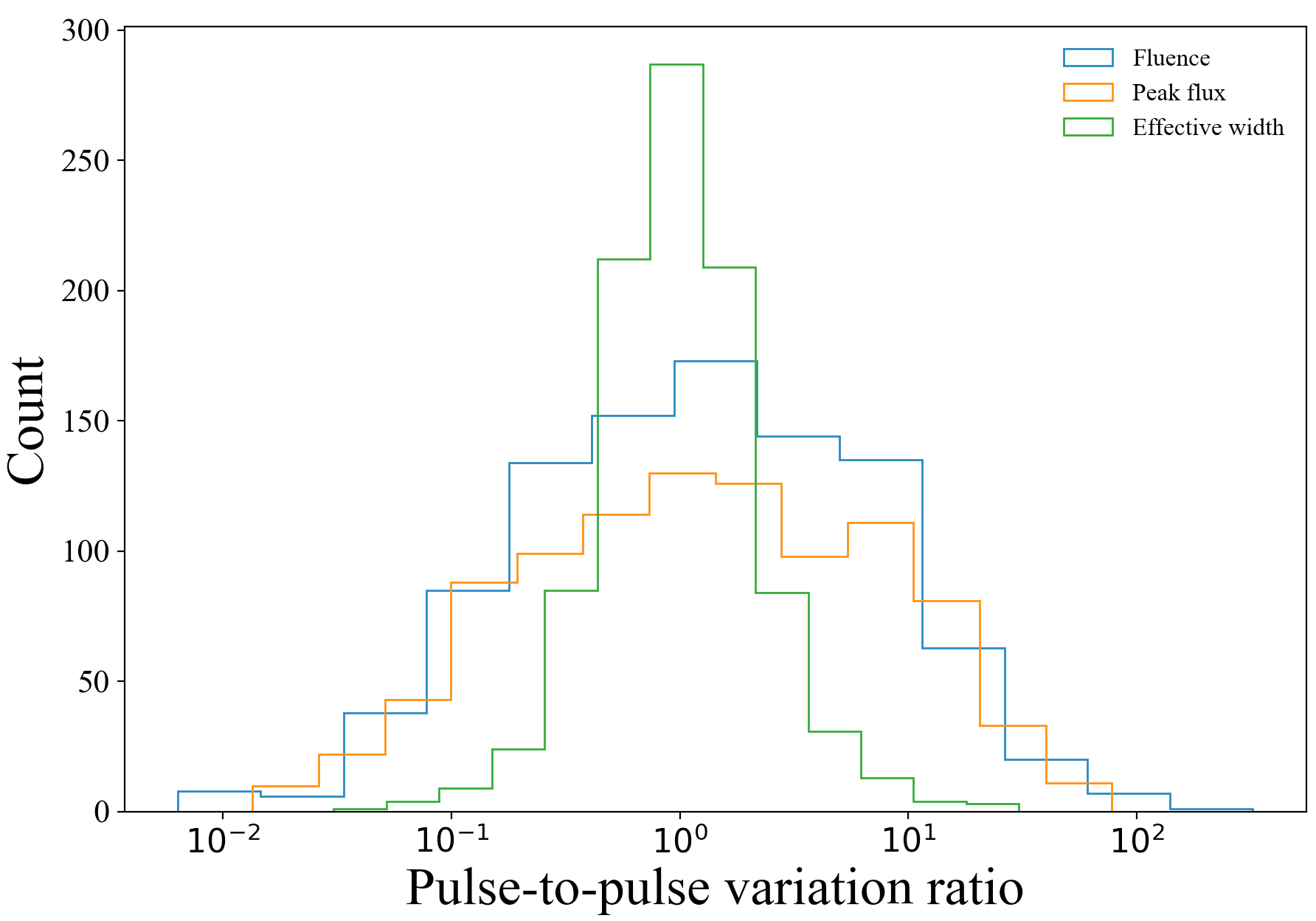}
  \caption{{\bf Pulse-to-pulse variation observed between adjacent pulses with waiting times of $1P_0$.} The distributions of ratios for fluence, peak flux, and effective width between 966 pulses and their corresponding adjacent pulses are displayed in cyan, orange, and grey, respectively. }
\label{Pulse-to-pulse_variation}
\end{figure}

To further quantify the pulse variation of J1913$+$1330, we selected a sample of 966 pulses that have an adjacent pulse with a waiting time of $1P_0$. 
We defined a pulse-to-pulse variation ratio as the ratio of fluence, peak flux, and effective width between the pulse and its immediate adjacent pulse. 
The distributions of these variation ratios are shown in Figure~\ref{Pulse-to-pulse_variation}. 
They can be well-fitted by log-normal distributions, with the mean value and standard deviation of log10($F$), log10($S_{\rm peak}$) and log10($W_{\rm ef}$) being 0.091 and 0.75, 0.099 and 0.76, -0.008 and 0.35, respectively. 
The narrow distribution of the effective width ratio indicates that the change in pulse width is relatively small. 
However, the fluence and the peak flux variation ratio distributions show that they can change by more than two orders of magnitude between adjacent pulses.

\subsection{RRAT~J1913$+$1330, pulsars and FRBs}

Thanks to the high sensitivity of FAST, we have been able to resolve the individual pulses in the so-called ``weak persistent emission mode'' of J1913$+$1330. However, in contrast to previous observations that showed two distinct emission modes~\citep{Bhattacharyya18}, our observations demonstrate that strong pulses are always detected together with the weak persistent emission mode within the pulse clusters.
The event rate of strong pulses during our observation is $\sim 11\,{\rm hr}^{-1}$, which is consistent with the results of other observations with shallower sensitivity~\citep{McLaughlin09, Bhattacharyya18}, and approximately one-twentieth of our detected rate of all pulses, which is $219\,{\rm hr}^{-1}$.
Around 49.4\% of the detected pulses were within $1P_0$ of another detected pulse, but the analysis of adjacent pulses shows changes in fluence and peak flux of more than two orders of magnitude. The broad intensity distribution and strong pulse-to-pulse variation of J1913+1330 clearly illustrate that observations with insufficient sensitivity can only detect sporadic pulses with a much lower event rate from such sources. 

\begin{figure}
\centering
\begin{tabular}{cc}
\includegraphics[width=80mm]{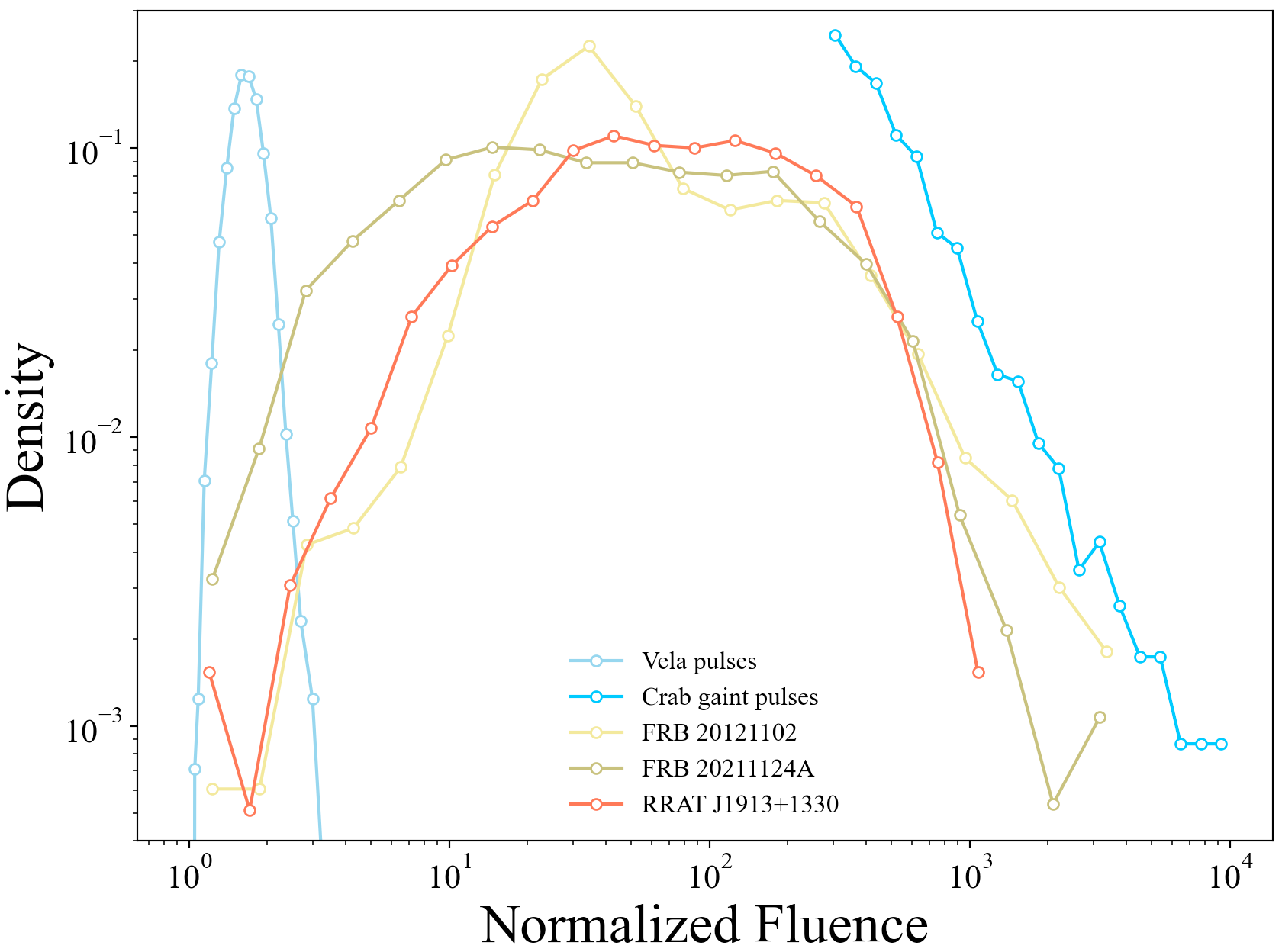}  
\end{tabular}
\caption{{\bf Normalized fluence distributions of the single pulses from RRAT~J1913$+$1330, bursts from the repeating FRB~20121102 and FRB~20201124A, giant pulses from the Crab pulsar, and single pulses from the Vela pulsar.} The fluence values of these sources are normalized to the faintest bursts, except for the Crab pulsar, which is normalized based on its mean fluence at 1400\,MHz. The density represents the number of bursts per bin divided by the total number of detected bursts.}
\label{flue_dis_4}
\end{figure}

In observations with sufficient sensitivity, J1913$+$1330 would exhibit an emission mode similar to the pulsars with nulling phases~\citep{Biggs92, Wang07, Burke-Spolaor12}. 
We have obtained numerous individual pulses of J1913$+$1330, revealing: (1) a gradual increase at low energy, (2) a distribution spanning three orders of magnitude, and (3) a power-law-like tail at high energy.
We compared the normalized fluence distribution of the single pulses from RRAT~J1913$+$1330, bursts from the repeating FRB~20121102~\citep{Li21} and FRB~20201124A~\citep{Xu22}, giant pulses from the Crab pulsar~\citep{Bera19}, and single pulses from one of the brightest pulsars, the Vela pulsar~\citep{Johnston01} in Figure~\ref{flue_dis_4}.
The power-law tail is similar to giant pulses.
Surprisingly, we found that the overall normalized distribution of J1913$+$1330 looks more like the distributions of repeating FRBs instead of pulsars. 
It is notable that strong pulse-to-pulse variations in fluence, pulse profile and polarization properties of J1913+1330 suggest an emission that is extremely variable. Such extreme variations in behaviour are common in the observations detecting large samples of bursts for repeating FRBs~\citep{Li21, Xu22}. 
These results suggest that J1913+1330 shares certain properties with populations of nulling pulsars, giant pulses and FRBs from different perspectives.

\subsection{An unstable sparking pulsar}
\begin{figure}%[tbp]
\centering
    \includegraphics[width=1\linewidth]{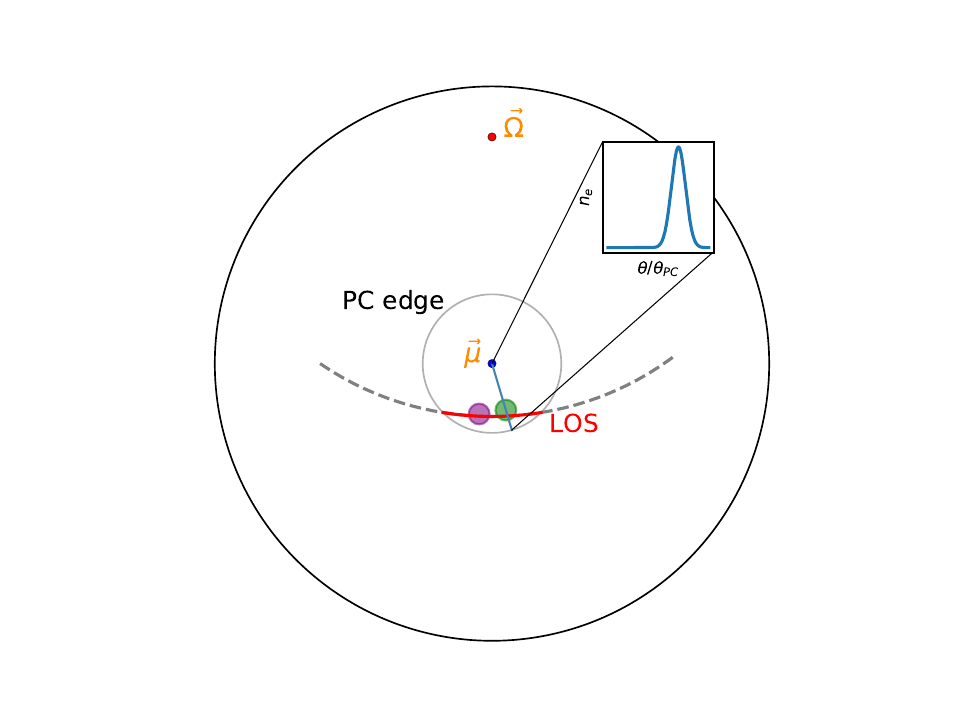}
    \caption{{\bf The schematic picture for the wandering spots scenario.}
    The polar cap region ($\theta \le \theta_{\rm PC}$) is marked by the inner grey circle as projected in the plane of the sky. 
    The symbols ``$\vec{\mu}$'' and ``$\vec{\Omega}$'' stand for the magnetic and rotation poles.
    The green and purple circles represent the discharging spots with a charge density of Gaussian distribution across the area. 
    The line of sight is shown by the red solid line and grey dashed line for the on and off phases respectively.
    }
    \label{fig:Schematic}
\end{figure}

\begin{figure}%[tbp]
\centering
    \includegraphics[width=1\linewidth]{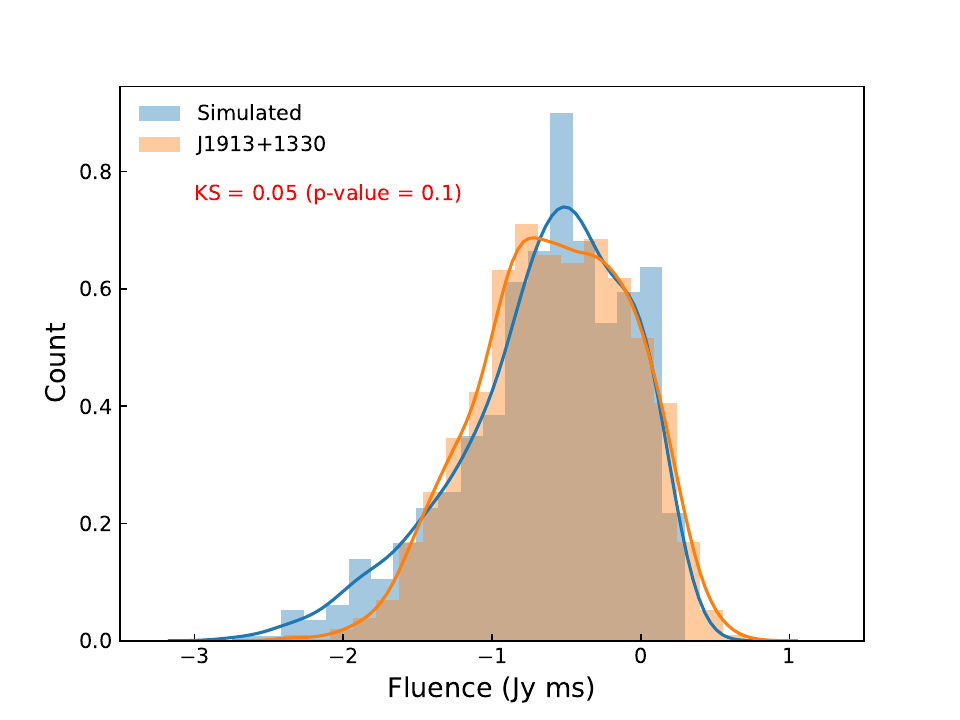}
    \caption{{\bf A comparison of simulated pulse energy distribution with that observed.}  
    The histogram of the observed and the simulated pulse fluence is shown by blue shadow and orange shadow respectively.
    The two-sample Kolmogorov-Smirnov test gives 0.05 with a p-value of 0.1.}
    \label{fig:Mock-ED}
\end{figure}

Coherent radio emissions from pulsars or RRATs are speculated to come from discharges of the polar cap (with an edge angle of $\theta_{\rm PC}$) named sparks, which inject ultrarelativistic pairs into the magnetosphere and curvature radiation is subsequently released~\citep{Ruderman75}.
The condition for the failure of pair production defines the pulsar radio emission ``death line'' in the $P - \dot{P}$ diagram of pulsars~\citep{Chen93, Zhang00}. When a pulsar is getting slightly below the emission death line, it may become occasionally active only if the pair production condition is temporarily satisfied, or nulling otherwise~\citep{Zhang07}. The pulse profile and peak flux vary abruptly from pulse to pulse, indicating the dramatic variation of the sparks and relevant emission processes in RRAT~J1913$+$1330.
We suppose that this old pulsar is in a transition state near its death line so that the pair production is unstable~\citep{Zhang00, Zhang07}.  
In such a state, the potential drop across the polar cap may be inhomogeneous~\citep{Leeuwen12} so that the accelerated pairs
would form several sub-beams in the flux tube spanned by open field lines~\citep{Gil00}. 
Even for a homogeneous primary pairs distribution left from the cap, the spatial distribution of the secondary pairs by the $\gamma-\boldsymbol{B}$ pair production turns into Gaussian-like along the latitude of field lines intersecting the star surface~\citep{Beskin22}.
Therefore, we assume that the polar cap of RRAT~J1913$+$1330 is populated with a number of equivalent two-dimensional sparking spots with a characteristic radius of $r_{\rm sp}$. 
The secondary pairs may release the emission coherently at a range of several tens to hundreds times of $R_{\rm NS}$, depending on where charged bunch trains
(named solitons) are formed. As the emission direction is along the tangent of the curved field line, the variation of the emission radius then leads to the oscillation of sparking spots along the latitudes to the line of sight.
On the other hand, the well-known circumferential $\boldsymbol{E} \times \boldsymbol{B}$ drift~\citep{Ruderman75, Deshpande01} or other local magnetospheric evolutions could make the spot oscillate in the azimuthal direction.
The line-of-sight cut on the emission beam of wandering spots would sample diverse pulse profiles, and hence account for the energy distribution of pulses of RRAT~J1913$+$1330 as shown below. 

The spatial distribution of the number density of secondary pairs from the $i$-th spot is assumed to be Gaussian-like centered at polar angles 
($\theta_{c,i}, \phi_{c,i}$) with width of $\sigma_i$ on the NS surface, i.e.,
\begin{equation}
n_0 \simeq \mathcal{M} n_{\rm GJ} f(\theta_{c,i},\phi_{c,i},\sigma_i) = \mathcal{M} n_{\rm GJ} \exp[-\mathcal{R}^2 / (2 \sigma_i^2)],
\label{Eq:density}
\end{equation}
where $n_{\rm GJ} \simeq \frac{B_{\rm s} \cos \alpha}{P_{\rm NS} q_e c}$ is the local Goldreich–Julian number density~\citep{Goldreich69},
$B_{\rm s}$ is the surface magnetic field, $P_{\rm NS}$ is the rotation period, $\alpha$ is the inclination angle between the magnetic pole and the spin axis, $\mathcal{M}$ is the multiplicity factor, and $\mathcal{R}$ is the distance from the spot center.
When the secondary pairs stream along the dipole magnetic field,
the radial dependence of the density is $n(r) \simeq n_0 (r/R_{\rm NS})^{-3}$.
For the $j$-th soliton moving with a field line tangent of $\hat{\bf{b}}$, the received radiation intensity by the observer in the direction of $\hat{\bf{k}}$ is $F_{j,\nu} (r,\gamma_0,\theta_{kb})$ (see Appendix for detailed formulation), where $\gamma_0$ is the bulk Lorentz factor of the pairs, $\theta_{kb}$ is the angle between the line of sight and the soliton's motion direction. At any instance, the observed flux density is the sum of the contribution of solitons formed along open field lines, i.e.,
\begin{equation}
F_{\nu} = \Sigma_j F_{j,\nu} (r,\gamma_0,\theta_{kb}).
\end{equation}

The stacked pulse profile of RRAT~J1913$+$1330 shows at least two main emission components. As a simplified but representative case, we adopt two wandering spots to mimic its pulse profile, in which the center of two Gaussian-like spots changes with each rotation of the NS (see Figure \ref{fig:Schematic}).
More specifically, $\theta_{c,i}$ is fixed while $\phi_{c,i}$ varies slightly but
randomly between each rotation, within a range of [$-\phi_{M}$, $-\phi_{m}$] and [$\phi_{m}$, $\phi_{M}$] respectively.
Due to a poor understanding of instabilities that drive the coherence,
the possibility that solitons are generated at $r_{\rm emi}$ by pairs injected from each spot is assumed to be uniform in a range of $R_{{\mathrm m},i} \le r_{\rm emi} \le R_{{\mathrm M},i}$.
In one spin period, we draw a random distribution of sub-sources in a radial range of $r_{\rm emi}$ to $r_{\rm emi} + \Delta r$, 
and density from Equation~\ref{Eq:density} to calculate one mock pulse profile.
We performed this calculation repeatedly and obtained a cluster of mock profiles, from which the simulated pulse energy distribution is derived.

Previous studies suggest that the coherent emission radius is roughly several tens of $R_{\rm NS}$ and the Lorentz factor of pairs is $\sim 10^2$ in normal pulsars~\citep{Mitra20}, which are used as priors here. The flux density level of RRAT~J1913$+$1330 requires that the multiplicity factor should be $\ge 10^4$, consistent with values given in theories \citep{Timokhin15}.
If we further assume that the magnetic inclination angle $\alpha$ is $30^{\circ}$~\citep{Wang23RAA}, then the maximum pulse width ($\sim 10$~ms, see Figure \ref{f_wid_flx}) gives an impact angle of $\beta \sim 36^{\circ}$ by considering that the maximum pulse width corresponds to the arc of the line-of-sight cutting on the polar cap region spanned at a height of $50~R_{\rm NS}$ \citep{Radhakrishnan69}.
On the other hand, the two misaligned log-normal components in the fit to the fluence distribution (Figure \ref{flue_dis}) hints that the relevant geometry of these two spots should be different, e.g., one is relatively wide, the other is narrow.

Due to the huge computational time costs to obtain one mock energy distribution, the Bayesian method is not adopted here to derive the best fitting parameters accurately. In practice, we adjust these parameters several times to reach an acceptable fitting goodness.
Using a wide spot of $\sigma_1 = 0.025~\theta_{\rm PC}$ and $r_{\mathrm{em},1} \in [42, 48]~R_{\rm NS}$, and a narrow spot of $\sigma_2 = 0.015~\theta_{\rm PC}$ and $r_{\mathrm{em},2} \in [52, 56]~R_{\rm NS}$, together with $\alpha = 30^{\circ}$, $\beta = 37^{\circ}$, $\Delta r = 1.5 R_{\rm NS}$, $\theta_{c,1} = \theta_{c,2} = 0.75~\theta_{\rm PC}$, $\phi_{m} = 0.015$ rad, $\phi_{M} = 0.1$ rad, $\gamma_0 = 200$, $\mathcal{M} = 10^4$, we got a simulated energy distribution in good agreement with that of RRAT~J1913$+$1330, with a Kolmogorov-Smirnov test value of 0.05 (Figure \ref{fig:Mock-ED}). Note that it is difficult to constrain these parameters strictly since we have limited observational constraints and a large number of free parameters, a common situation in the literature. Our calculation provides one set of reasonable parameters, yet other choices may give a better-fitting goodness.

%%%%%%%%%%%%%%%%%%%%%%%%%%%%%%%%%%%%%%%%%%%%%%%%%%%%%%%%%%%%%%%%%%%%%%%%%%%%%%%%
\section{Conclusion} \label{sec:con}
%%%%%%%%%%%%%%%%%%%%%%%%%%%%%%%%%%%%%%%%%%%%%%%%%%%%%%%%%%%%%%%%%%%%%%%%%%%%%%%%
RRATs like J1913$+$1330 are known to exhibit sporadic bursts in observations with insufficient sensitivity~\citep{McLaughlin09, Bhattacharyya18}. 
However, in our observation of RRAT~J1913$+$1330 using the highly sensitive FAST telescope, we found that its emission consists of clustered and sequential pulses, with significant
variations of more than two orders of magnitude between adjacent sequential pulses. 
The energy distribution of the detected individual pulses exhibited a range spanning three orders of magnitude. 
Our results offer further evidence that RRATs can be explained as pulsars with extreme pulse-to-pulse modulation~\citep{Weltevrede06}. 
%Future observation
This emission pattern could be phenomenologically understood by the cut of the line of sight on the emission beams from wandering sparking spots above the polar cap region. Further efforts are invoked to unveil the physics that drives the sparking fluctuation. 
The presence of sequential pulse trains during active phases, along with significant pulse variations in profile, fluence, flux, and width, should be intrinsic to a subset of RRATs. However, radio telescopes with limited sensitivity can only detect brighter and infrequent pulses from this kind of RRAT.

Our analysis of the properties of individual pulses from J1913$+$1330 indicates that J1913+1330 represents a peculiar source that shares certain properties with populations of nulling pulsars, giant pulses and FRBs from different perspectives.
A sample of RRATs exhibiting similar properties to J1913+1330 has the potential to significantly enhance our understanding of pulsars, RRATs, and FRBs.
Therefore, further detailed classification and monitoring of RRATs is encouraged to determine if they share similar properties with J1913$+$1330. Telescopes with large collection areas (such as SKA or the new proposed FAST array) or wide bandwidths have the potential to unveil the ``true energy distribution'' of more RRATs by detecting their ``weak pulses''.

%% Also note that the akcnowlodgment environment does not support long amounts of text. If you have a lot of people and institutions to acknowledge, do not use this command. Instead, create a new 
\section*{Acknowledgments}
We would like to express our gratitude to Apurba Bera for providing the Crab pulsar data, which was invaluable for our research. This work is partially supported by the National SKA Program of China (2022SKA0130100), the National Natural Science Foundation of China (grant Nos. 12041306, 12273113,12233002,12003028), the international Partnership Program of Chinese Academy of Sciences for Grand Challenges (114332KYSB20210018), the National Key R\&D Program of China (2021YFA0718500), the ACAMAR Postdoctoral Fellow, China Postdoctoral Science Foundation (grant No. 2020M681758), and the Natural Science Foundation of Jiangsu Province (grant Nos. BK20210998). JJG acknowledges the support from the Youth Innovation Promotion Association (2023331). JSW acknowledges the support from the Alexander von Humboldt Foundation. YFH acknowledges the support from the Xinjiang Tianchi Program.\\

\appendix
\section{Formulation of soliton radiation}
Recent kinetic plasma simulations and theoretical studies show growing support on the existence of long-lasting charged solitons that provide coherent radio emission.
Considering solitons emerging from an initial white noise field with the dimensionless correlation wavenumber ($k_c^{\prime}$) 
in the plasma frame, and $N_b$ bunches within each soliton, 
the corresponding wavenumber of the bunch in the observer frame is~\citep{Rahaman22} 
\begin{equation}
k_0 = 6.7 \times 10^{-2} \left( \frac{N_b}{10} \right) \left(\frac{\gamma^{\prime}}{3}\right)^{1/2} \left(\frac{\gamma_0}{200} \right) 
\left(\frac{k_c^{\prime}}{2}\right) \mathrm{cm}^{-1}, 
\end{equation}
and its longitudinal length is $s_0 = 2 \pi /k_0$,
where $\gamma_0$ is the bulk Lorentz factor of the pairs and $\gamma^{\prime}$ is average Lorentz factor in the plasma frame.
For a perturbation wave with a phase velocity of $v_p$ ($\beta_p = v_p/c$) in the plasma frame, the observed flux density from each bunch could be written as~\citep{Buschauer76}
\begin{equation}
F_{b,\nu} = \frac{36^{2/3}}{16 \pi^2 c D_L^2} Q_{b}^2 \omega^{2/3} \left(\frac{c}{R_c}\right)^{1/3} 
\left[ \frac{\sin(k \frac{\eta_0}{2} \theta_{kb})}{k \frac{\eta_0}{2} \theta_{kb}} \right]^4 
\left[ \frac{\sin((k_0-k)\frac{s_0}{2})}{(k_0-k)\frac{s_0}{2}} \right]^2 
\left| L_2 (z) \right|^2, 
\label{Eq:F-bunch}
\end{equation}
where $Q_b = q_e n(r) s_0 \eta_0^2$ is the charge number of the bunch with
a transverse length of $\eta_0 \simeq (R_c/6k^2)^{1/3}$, $R_c$ is the curvature radius, $\theta_{kb}$ is the angle between the line of sight ($\hat{\bf{k}}$) and the tangent of the local field ($\hat{\bf{b}}$),
and $L_2$ is related to Bessel functions that determine the radiation pattern, i.e.,
\begin{eqnarray}
z & = & \frac{1}{2} \left( \frac{6 \omega^2 R_c^2}{c^2} \right)^{1/3} \left\{ \frac{1}{\gamma_0^2} 
\left[1- 2 h(\beta_p) \frac{k_{0}}{k} \right] + \theta_{kb}^2 \right\},\\
h(\beta_p) &=& \frac{\beta_p}{1 + \frac{v_0}{c} \beta_p} \le \frac{1}{2},\\
L_2 (\left|x\right|) & = & i \frac{2}{3^{3/2}} \left|x\right| K_{2/3} \left[ \frac{2}{3^{3/2}} \left|x\right|^{3/2} \right], \\
L_2 (-\left|x\right|) & = & i \frac{2 \pi}{9} \left|x\right| \left\{ J_{-2/3} \left[ \frac{2}{3^{3/2}} \left|x\right|^{3/2} \right] - 
J_{2/3} \left[ \frac{2}{3^{3/2}} \left|x\right|^{3/2} \right] \right\}.
\end{eqnarray}
Finally, the observed flux from the $j$-th soliton is $F_{j,\nu} = N_b F_{b,\nu}$.
In our calculations, we choose a $k_c^{\prime} = 7.86$ to select the wavenumbers being radiated intensely around $k \approx k_0$,
which means faint solitons with moderate deviations from $k$ are omitted.
Also, typical values of $N_b = 10$ and $\gamma^{\prime} = 3$ are adopted for all solitons.


\bibliography{J1913}{}
\bibliographystyle{aasjournal}



\end{document}